\documentclass[%
aip,
amsmath, amssymb,
preprint,%
]{revtex4-1}

\usepackage{graphicx, color}
\usepackage{dcolumn}
\usepackage{bm}

\usepackage[utf8]{inputenc}
\usepackage[T1]{fontenc}
\usepackage{mathptmx}
\usepackage{etoolbox}
\usepackage{comment}
\usepackage[subrefformat=parens]{subcaption}
\usepackage[labelsep=period, figurename=Fig.\ , tablename=Table\ , singlelinecheck=false]{caption}
\usepackage{bm}
\usepackage{amsmath}
\usepackage{amsthm}

\newcommand{\one}{I}
\newcommand{\two}{I\hspace{-1.2pt}I}
\newcommand{\three}{I\hspace{-1.2pt}I\hspace{-1.2pt}I}

\theoremstyle{definition}
\newtheorem{ass}{Assumption}

\makeatletter
\def\@email#1#2{%
 \endgroup
 \patchcmd{\titleblock@produce}
  {\frontmatter@RRAPformat}
  {\frontmatter@RRAPformat{\produce@RRAP{*#1\href{mailto:#2}{#2}}}\frontmatter@RRAPformat}
  {}{}
}%
\makeatother
\begin{document}

\preprint{AIP/123-QED}

\title{Bifurcation analysis of a two-neuron central pattern generator model for both oscillatory and convergent neuronal activities}

\author{Kotaro Muramatsu}
\affiliation{ 
Department of Mathematical Informatics, Graduate School of Information Science and Technology, The University of Tokyo, Japan
}%

\author{Hiroshi Kori}%
\affiliation{ 
Department of Mathematical Informatics, Graduate School of Information Science and Technology, The University of Tokyo, Japan
}%
\altaffiliation[Also at ]{Department of Mathematical Informatics, Graduate School of Information Science and Technology, The University of Tokyo, Japan}
 
 \email{oratok-m@g.ecc.u-tokyo.ac.jp, kori@k.u-tokyo.ac.jp}

\homepage{}
\affiliation{%
Department of Complexity Science and Engineering, Graduate School of Frontier Sciences, The University of Tokyo, Japan
}%


\begin{abstract}
The neural oscillator model proposed by Matsuoka is a piecewise affine system, which exhibits distinctive periodic solutions. Although such typical oscillation patterns have been widely studied, little is understood about the dynamics of convergence to certain fixed points and bifurcations between the periodic orbits and fixed points in this model. We performed fixed point analysis on a two-neuron version of the Matsuoka oscillator model, the result of which explains the mechanism of oscillation and the discontinuity-induced bifurcations such as subcritical/supercritical Hopf-like, homoclinic-like, and grazing bifurcations. Furthermore, it provided theoretical predictions concerning a logarithmic oscillation-period scaling law and noise-induced oscillations, which are both observed around those bifurcations. These results are expected to underpin further investigations into both oscillatory and transient neuronal activities with respect to central pattern generators.
\end{abstract}

\maketitle

\begin{quotation}
The Matsuoka oscillator model is a neuronal network model, which exhibits oscillatory activities owing to the adaptation property of each neuron and the mutual inhibitions between neurons. This is often applied to modeling the spinal oscillatory neuronal circuits known as central pattern generators (CPGs) and for simulating biological locomotion such as human bipedal walking. However, most previous studies have overlooked its dynamics that converge toward stationary states, corresponding to transient neuronal activities and non-oscillatory movements. In this study, we conducted fixed point analysis on a two-neuron case of the Matsuoka oscillator model. We (\one) formulated the existence and stability of all possible fixed points, (\two) demonstrated the emergence of oscillatory solutions and bifurcation mechanisms between oscillatory and convergent dynamics, and (\three) predicted a logarithmic oscillation-period scaling law and noise-induced oscillation. Our results indicate that central nervous systems might take advantage of CPGs for rhythmic locomotion and non-oscillatory or discrete movements. The discussion of limitations presented herein will, in the future, probably be followed by extending the Matsuoka oscillator model to understand an integrative mechanism for neural control of both rhythmic and discrete movements.
\end{quotation}

\section{INTRODUCTION \label{sec1}}

A biological neural circuit, CPG, is the basis of rhythmic movements, e.g., locomotion and respiration, in animals.\cite{ijspeert2007swimming, ijspeert2008central} The CPGs of vertebrates are located in the spinal cord. Furthermore, CPGs can generate stable oscillatory activities by receiving stationary inputs or tonic drives descending from a part of the brainstem, the mesencephalic locomotor region (MLR). The CPGs oscillation patterns are also modulated by feedback from peripheral sensory organs, contributing to highly autonomous and adaptive motor control. Recent studies have suggested that the spinal interneuronal networks thought to implement CPGs are also involved in discrete, transient, non-oscillatory movements such as point-to-point reaching with the upper limbs.\cite{grillner2006biological, takei2013spinal, azim2014skilled}

Various dynamical system models mathematically describe the CPG functions. The Matsuoka oscillator model, which focuses on the dynamics of neuronal firing rates, is an example.\cite{matsuoka1985sustained, matsuoka1987mechanisms, matsuoka2011analysis} Despite having less precise temporal resolution than spiking neuronal network models,\cite{angelidis2021spiking} firing rate models are useful for understanding neural phenomena in macroscopic timescales such as neuromuscular activity and motor control. Thus, the Matsuoka oscillator model has been applied to the simulation studies of human bipedal walking.\cite{taga1991self, taga1995model, ogihara2001generation} Additionally, the physiological interpretation of the model is feasible, compared to more abstract models without neuronal configuration such as the phase oscillator model.\cite{ijspeert2007swimming} The Wilson-Cowan model\cite{wilson1972excitatory, wilson1973mathematical} is another firing rate model, although it addresses neither the adaptation properties nor the mutual inhibitory connections of neurons. In particular, the latter feature follows the half-center structure of spinal circuits,\cite{ijspeert2008central} which validates the Matsuoka oscillator as a basic hypothetical model for CPGs.

For dynamical systems, rhythmic oscillatory activities are represented as stable limit cycles in the Matsuoka oscillator model. The original paper on this model (Ref.~\onlinecite{matsuoka1985sustained}) presents several proofs for the existence condition of oscillatory solutions. However, previous studies have mainly focused only on its distinctive oscillatory solutions,\cite{degallier2010modeling} whereas some transient activities with convergence to stationary states or fixed points have been insufficiently discussed. Ref.~\onlinecite{de2003interaction} highlighted this problem; however, the terminal stationary states are not realized by positive constant inputs and positive neuronal inner states. Furthermore, little is known and understood on the bifurcation types, mechanisms, and critical behaviors in the vicinity of the bifurcations that emerge between such fixed points and the well-known limit cycles. Therefore, further investigation into attracting fixed points and their relevant bifurcations is essential for grasping the possibility of transient activities in the Matsuoka oscillator model, which could be related to discrete, transient movements.

This study mainly aims to systematically formulate possible fixed points corresponding to stationary states that appear in the Matsuoka oscillator model and to investigate bifurcations between oscillatory solutions and fixed points in addition to the oscillation mechanism. We first review the original Matsuoka oscillator model (Section~\ref{sec2}); thereafter, we discuss the classification of oscillation types and approximations of the oscillation period (Section~\ref{sec3}). This review is followed by a comprehensive fixed point analysis regarding the existence and stability of fixed points (Section~\ref{sec4}). Based on the analysis results, we present theoretical formulations to explain the emergence of oscillations and bifurcation scenarios (Section~\ref{sec5}). A logarithmic oscillation-period scaling law and a novel prediction of noise-induced oscillations are also proposed. Throughout this paper, we attempt to provide a foundation for understanding the common mechanisms underlying the oscillatory and convergent neuronal activities that both emerge in one of the well-known CPG models.

\section{BASIS OF THE STUDY \label{sec2}}
\subsection{The Matsuoka Oscillator Model \label{subsec2A}}

The Matsuoka oscillator model\cite{matsuoka1985sustained, matsuoka1987mechanisms, matsuoka2011analysis} is a neural oscillator model comprising $n$ firing neurons with neuronal adaptation properties and mutual inhibitions. The time evolution of the $i$-th neuron (the neuron-$i$) is described using the following differential equations:
\begin{subequations}
\begin{eqnarray}
\tau_x \frac{dx_i}{dt} &=& -x_i-by_i - \sum_{j \neq i}^{n} a_{ij} z_j + s_i, \label{eq-1a}\\
\tau_y \frac{dy_i}{dt} &=& -y_i + z_i, \label{eq-1b}\\
z_i &=& \max(x_i, 0), \label{eq-1c}
\end{eqnarray}\label{eq-1}
\end{subequations}
where $x_i$ is the membrane potential or inner state of the neuron-$i$; $y_i$ is the variable of adaptation or fatigue; $z_i$ is the firing rate; $s_i$ is the constant input stimulus into the neuron-$i$; $a_{ij} \geq 0$ is the synaptic weight from neuron-$j$ to the neuron-$i$; $b>0$ is the constant determining adaptation intensity; $\tau_x>0$ and $\tau_y>0$ are the time constants of $x_i$ and $y_i$, respectively. Here, $a_{ij}$ is non-negative because the model supposes mutual inhibitions between neurons. No excitatory synapses or self-inhibitions are considered in this original form of the Matsuoka oscillator model. Although not specified in this study, the second is assumed as the time unit in most previous studies.

The outline of this model resembles regular recurrent neural networks (RNNs), adding the specific property of adaptation. The nonlinear transformation \eqref{eq-1c} is the same as that of the rectified linear unit (ReLU) function, which is piecewise linear as follows:
\begin{eqnarray}
z_i=
\left\{
\begin{array}{ll}
x_i & (x_i>0) \\
& \\
0 & (x_i \leq 0)
\end{array}
\right. . \label{z_i}
\end{eqnarray}
Note that this activation function should have a firing threshold $\theta$ as $z_i=\max(x_i-\theta, 0)$. This threshold parameter can be, however, erased without loss of generality by redefining $x_i-\theta$ and $s_i-\theta$ as $x_i$ and $s_i$, respectively.\cite{matsuoka1985sustained} By applying this procedure, the threshold parameter $\theta$ can be ignored in this study.

Despite being named an ``oscillator,'' a single neuron in the Matsuoka oscillator model ($n=1$) cannot sustainably oscillate by itself. According to the case classification \eqref{z_i}, a single neuron follows different dynamics depending on the positivity or negativity of $x_i$. For $ x_i \leq 0$, the system \eqref{eq-1} is rewritten as
\begin{subequations}
\begin{eqnarray}
\frac{d}{dt}
\begin{bmatrix}
x_i \\
y_i
\end{bmatrix}
=
\begin{bmatrix}
-\frac{1}{\tau_x} & -\frac{b}{\tau_x} \\
&\\
0 & -\frac{1}{\tau_y}
\end{bmatrix}
\begin{bmatrix}
x_i \\
y_i
\end{bmatrix}
+
\begin{bmatrix}
\frac{s_i}{\tau_x} \\
\\
0
\end{bmatrix}. \label{eq-signle-a}
\end{eqnarray}
Similarly, in the case $x_i>0$,
\begin{eqnarray}
\frac{d}{dt}
\begin{bmatrix}
x_i \\
y_i
\end{bmatrix}
=
\begin{bmatrix}
-\frac{1}{\tau_x} & -\frac{b}{\tau_x} \\
& \\
\frac{1}{\tau_y} & -\frac{1}{\tau_y}
\end{bmatrix}
\begin{bmatrix}
x_i \\
y_i
\end{bmatrix}
+
\begin{bmatrix}
\frac{s_i}{\tau_x} \\
\\
0
\end{bmatrix}. \label{eq-signle-b}
\end{eqnarray}
\end{subequations}
In both cases, the single-neuron state $\begin{bmatrix}x_i & y_i \end{bmatrix}^\top$ asymptotically converges to stable fixed points. For $x_i>0$, based on experimental knowledge, Ref.~\onlinecite{matsuoka1985sustained} introduced biological restriction for non-damped oscillation in the neuronal activities as follows:
\begin{eqnarray}
\frac{(\tau_y-\tau_x)^2}{4\tau_x\tau_y} \geq b, \label{eq-2}
\end{eqnarray}
The adaptation process requires $\tau_y>\tau_x$, meaning that $y_i$ is a slower variable than $x_i$. Additionally, Ineq.~\eqref{eq-2} requires even sufficient separation in the timescale between $\tau_y$ and $\tau_x$. No oscillatory solution is possible in the case $n=1$ of the Matsuoka oscillator model under these conditions for any input stimulus $s_i$, as shown in Fig.~1 in Ref.~\onlinecite{matsuoka1985sustained}.

\begin{figure*}
 \includegraphics[width=\linewidth]{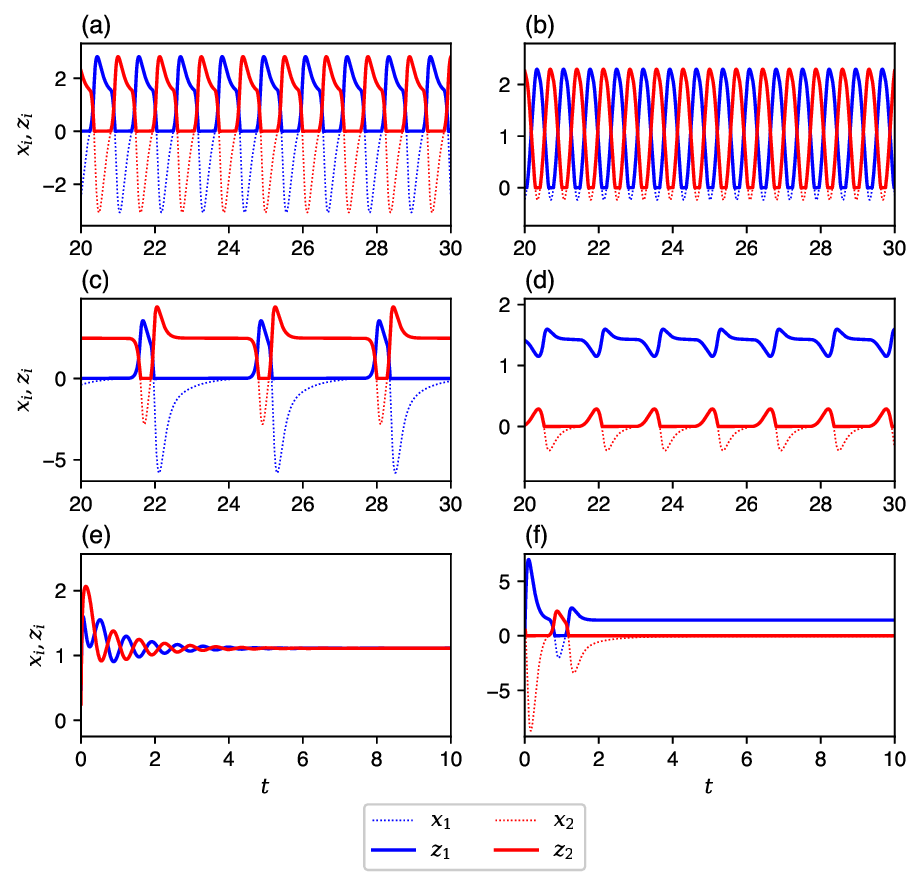}
 \caption{Neuronal activity patterns for $n=2$ of the Matsuoka oscillator model. Note that $a=a_{12}=a_{21}$, $r=\frac{s_2}{s_1}$, and the parameters $b$, $\tau_x$, $\tau_y$, and $s_1$ are all fixed as written in Subsection~\ref{subsec3A}. (a) $a=2$, $r=1$. (b) $a=2$, $r=1.73$. (c) $a=1.13$, $r=1$. (d) $a=1.6$, $r=0.47$. (e) $a=1$, $r=1$. (f) $a=2$, $r=0.56$.}
 \label{fig-t-xyz}
\end{figure*}

When mutually connected with inhibitory synapses, however, neurons of the Matsuoka oscillator model assume oscillatory properties as a network. In cases where more than three neurons are connected to each other ($n \geq 3$), many network topologies and corresponding diverse oscillation patterns can be observed as in Figs.~3--5 of Ref.~\onlinecite{matsuoka1985sustained}. The following sections report on the case of $n=2$ because the two-neuron model is the fundamental and simplest element of oscillatory circuits. This four-dimension system is written as
\begin{subequations}
\begin{eqnarray}
\tau_x \frac{dx_1}{dt} &=& -x_1-by_1 - a_{12} z_2 + s_1, \label{eq-3a}\\
\tau_y \frac{dy_1}{dt} &=& -y_1 + z_1, \label{eq-3b}\\
\tau_x \frac{dx_2}{dt} &=& -x_2-by_2 - a_{21} z_1 + r s_1, \label{eq-3c}\\
\tau_y \frac{dy_2}{dt} &=& -y_2 + z_2, \label{eq-3d}\\
z_1 &=& \max(x_1, 0), \label{eq-3e}\\
z_2 &=& \max(x_2, 0). \label{eq-3f}
\end{eqnarray}\label{eq-3}
\end{subequations}
Here, we define a new parameter $r$ as the input ratio between the two input stimuli
\begin{eqnarray}
r:=\frac{s_2}{s_1}. \label{eq-r}
\end{eqnarray}

Figure~\ref{fig-t-xyz} shows several neuronal activity patterns as solutions of the two-neuron model \eqref{eq-3} with symmetric connection $a=a_{12}=a_{21}$ and various values of $r$ and $a$. Although the system does not always have such oscillatory properties [e.g. Fig.~\ref{fig-t-xyz}(e), (f)], the observed oscillatory solutions are empirically stable limit cycles when it has them [e.g. Fig.~\ref{fig-t-xyz}(a)--(d)].

\subsection{Existence Condition of Oscillatory Solutions \label{subsec2B}}

The original papers (Ref.~\onlinecite{matsuoka1985sustained, matsuoka1987mechanisms}) derived the condition for the existence of oscillatory solutions concerning synaptic connection $a_{12}$, $a_{21}$, and input ratio $r$. When $n=2$, this condition is
\begin{subequations}
\begin{eqnarray}
\sqrt{a_{12}a_{21}} &>& a_{\mathrm{inf}}, \label{eq-5a}\\
r &>& r_{\mathrm{inf}}, \label{eq-5b}\\
r &<& r_{\mathrm{sup}}, \label{eq-5c}
\end{eqnarray}\label{eq-5}
\end{subequations}
where
\begin{subequations}
\begin{eqnarray}
a_{\mathrm{inf}} &:=& 1+\frac{\tau_x}{\tau_y}, \label{eq-a_inf}\\
r_{\mathrm{inf}} &:=& \frac{a_{21}}{1+b}, \label{eq-r_inf}\\
r_{\mathrm{sup}} &:=& \frac{1+b}{a_{12}}. \label{eq-r_sup}
\end{eqnarray}\label{inf_sup}
\end{subequations}
Ineq.~\eqref{eq-5a} requires that synaptic weights $a_{12}$ and $a_{21}$ are sufficiently large. Ineqs.~\eqref{eq-5b} and \eqref{eq-5c} provide a constraint for $r=\frac{s_2}{s_1}$ so that oscillatory solutions emerge when the level of input $s_2$ into the neuron-2 is roughly comparable with input $s_1$ into the neuron-1. From Ineqs.~\eqref{eq-5a}--\eqref{eq-5c}, a necessary condition for $a_{12}$ and $a_{21}$ is given as
\begin{eqnarray}
1+\frac{\tau_x}{\tau_y} < \sqrt{a_{12}a_{21}} < 1+b. \label{eq-add1}
\end{eqnarray}
Considering Ineqs.~\eqref{eq-2} and \eqref{eq-add1} together, we can obtain inequalities
\begin{eqnarray}
\frac{\tau_x}{\tau_y}  < b \leq \frac{(\tau_y-\tau_x)^2}{4\tau_x\tau_y}, \label{eq-add2}
\end{eqnarray}
which forms an additional necessary condition for $\tau_x, \tau_y>0$ as
\begin{eqnarray}
\tau_y > 3 \tau_x. \label{eq-add3}
\end{eqnarray}
Precisely, e.g., the parameter setting in Ref.~\onlinecite{matsuoka2011analysis} does not satisfy Ineq.~\eqref{eq-add3}, which inappropriately results in permitting the damped oscillation of a single neuron.

Ineqs.~\eqref{eq-5a}--\eqref{eq-5c} are the conditions under which there is no ``stable'' fixed point in the system. According to Ref.~\onlinecite{matsuoka1985sustained}, any solutions of the general system \eqref{eq-1a}--\eqref{eq-1c} are proved to be bounded for $t\geq0$. Hence, if Ineqs.~\eqref{eq-5a}--\eqref{eq-5c} are satisfied, the system state does not converge to any fixed points but only travels in the bounded region; thus, nonstationary solutions occur. Note that these oscillations are empirically observed as limit cycles, which has not been mathematically guaranteed yet.

For the symmetric network $a=a_{12}=a_{21}$, we can rewrite Ineqs.~\eqref{eq-5a}--\eqref{eq-5c} as

\begin{subequations}
\begin{eqnarray}
a&>& a_{\mathrm{inf}}, \label{eq-4a}\\
r &>& r_{\mathrm{inf}}, \label{eq-4b}\\
r &<& r_{\mathrm{sup}}, \label{eq-4c}
\end{eqnarray}\label{eq-4}
\end{subequations}
where
\begin{subequations}
\begin{eqnarray}
a_{\mathrm{inf}} &:=& 1+\frac{\tau_x}{\tau_y}, \label{eq-a_inf_sym}\\
r_{\mathrm{inf}} &:=& \frac{a}{1+b}, \label{eq-r_inf_sym}\\
r_{\mathrm{sup}} &:=& \frac{1+b}{a}. \label{eq-r_sup_sym}
\end{eqnarray}\label{eq-6}
\end{subequations}

\section{NUMERICAL OBSERVATIONS \label{sec3}}

\subsection{Classification of Oscillation Types \label{subsec3A}}

\begin{figure}
 \includegraphics[width=\linewidth]{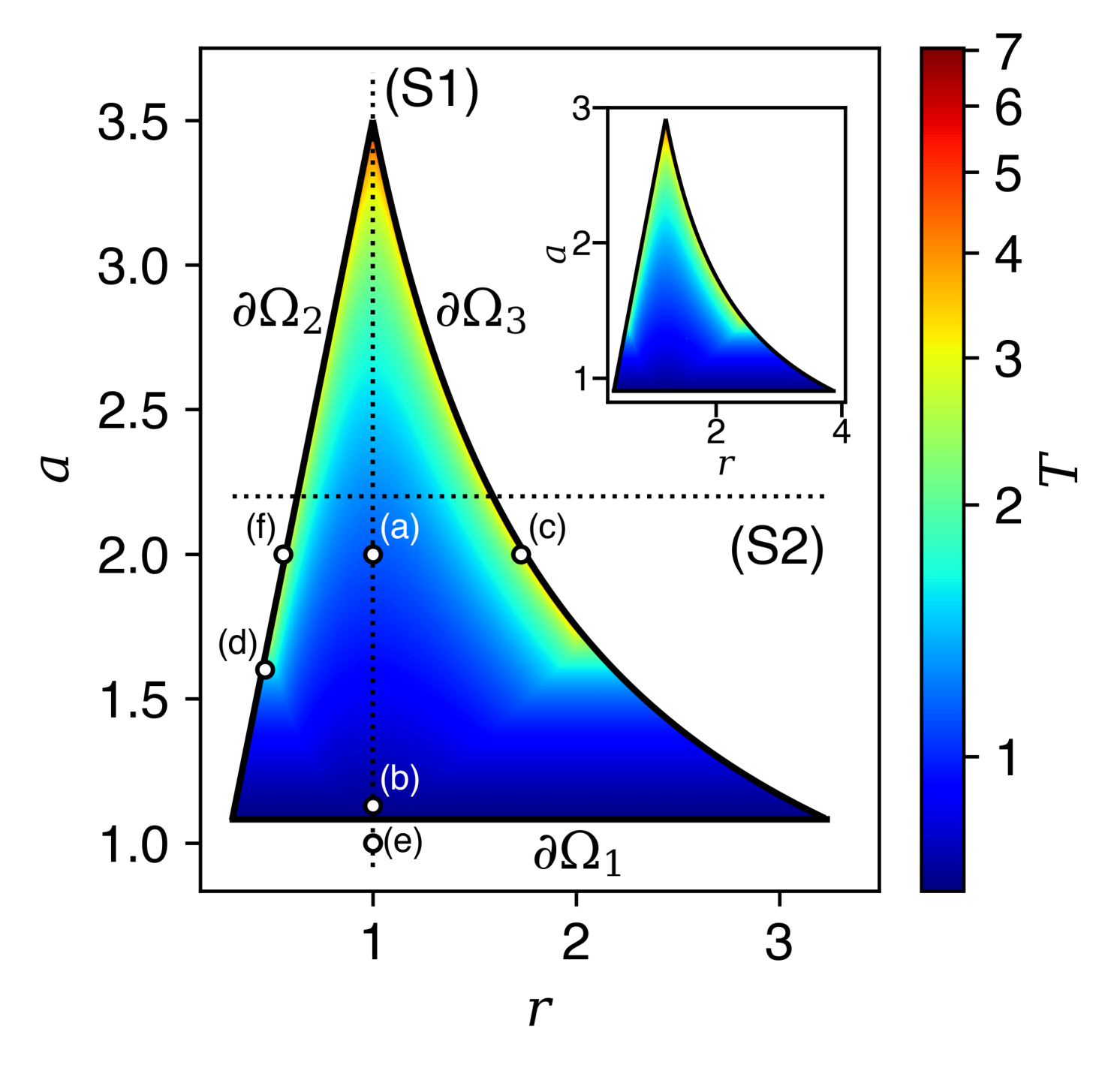}
 \caption{Plot of oscillation period $T$ of limit cycle solutions on the $r$-$a$ plane in the symmetric case $a=a_{12}=a_{21}$ of the system \eqref{eq-3}. Oscillatory solutions exist only in $\Omega$ ($T$ is plotted), which is surrounded by the three borderlines $\partial \Omega_1$, $\partial \Omega_2$, and $\partial \Omega_3$ (black solid lines). Each circle plot indicates the parameter set $(r,a)$ corresponding to the activity patterns of Fig.~\ref{fig-t-xyz}(a)--(f). (Inset) the similar plot in the asymmetric connection case where $a=a_{12}$ and $a_{21}=1.44 a$.}
 \label{fig-p}
\end{figure}

Because we are interested in the effect of the neuronal interactions and external inputs, we do not consider the changes in values of $b$, $\tau_x$, $\tau_y$. Moreover, the multiplication of $s_1$ and $s_2$ by a common positive constant $\mu$ only increases the amplitude $\mu$ times with no change in the period or frequency of oscillation owing to the piecewise linearity of the system.\cite{matsuoka1987mechanisms} Although $s_1$ is fixed, changing input ratio $r=\frac{s_2}{s_1}$ introduces an asymmetry between the external inputs into two neurons, which qualitatively transforms the system dynamics. Therefore, in the following discussions on the bifurcations of this model, we mainly attempt to vary parameters $r$, $a_{12}$, and $a_{21}$ for fixed $\tau_x$, $\tau_y$, $b$, and $s_1$ values. Additionally, in all numerical simulations and figures in this study, we fix $\tau_x = 0.05$, $\tau_y = 0.6$, $b = 2.5$, and $s_1=5$. Note that the positivity or negativity of $s_1$ is still important concerning the condition of existence for several fixed points, as discussed in the next section.

Figure~\ref{fig-p} shows the oscillation period in the $(r,a)$ space for the symmetric network $a=a_{12}=a_{21}$. On this plane, three critical borderlines are defined by
\begin{subequations}
\begin{eqnarray}
\partial \Omega_1 &:=& \left\{ (r,a) \in \mathbb{R} \times \mathbb{R}_{>0} \mid a = a_{\mathrm{inf}} \right\}, \label{eq-6a} \\
\partial \Omega_2 &:=& \left\{ (r,a) \in \mathbb{R} \times \mathbb{R}_{>0} \mid r = r_{\mathrm{inf}} \right\}, \label{eq-6b} \\
\partial \Omega_3 &:=& \left\{ (r,a) \in \mathbb{R} \times \mathbb{R}_{>0} \mid r = r_{\mathrm{sup}} \right\}, \label{eq-6c}
\end{eqnarray}\label{eq-6}
\end{subequations}
where $a_{\mathrm{inf}}$, $r_{\mathrm{inf}}$, and $r_{\mathrm{sup}}$ are given by Eqs.~\eqref{eq-a_inf_sym}--\eqref{eq-r_sup_sym}, respectively. Specifically, we can obtain a certain borderline threshold $a_{+}^*$, which meets $a_{\mathrm{inf}} < a_{+}^*<1+b$, and can define the upper and lower parts of $\partial \Omega_2$ and $\partial \Omega_3$:
\begin{subequations}
\begin{eqnarray}
\partial \Omega_2^+ &:=& \left\{ (r,a) \in \mathbb{R} \times \mathbb{R}_{>0} \mid r = r_{\mathrm{inf}},  a \geq a_+^*  \right\}, \label{eq-omegapm-a} \\
\partial \Omega_2^{-} &:=& \left\{ (r,a) \in \mathbb{R} \times \mathbb{R}_{>0} \mid r = r_{\mathrm{inf}}, a < a_+^* \right\}, \label{eq-omegapm-b} \\
\partial \Omega_3^+ &:=& \left\{ (r,a) \in \mathbb{R} \times \mathbb{R}_{>0} \mid r = r_{\mathrm{sup}}, a \geq a_+^* \right\}, \label{eq-omegapm-c} \\
\partial \Omega_3^{-} &:=& \left\{ (r,a) \in \mathbb{R} \times \mathbb{R}_{>0} \mid r = r_{\mathrm{sup}}, a < a_+^* \right\}. \label{eq-omegapm-d}
\end{eqnarray}\label{eq-omegapm}
\end{subequations}
The borderlines $\partial \Omega_1$, $\partial \Omega_2$, and $\partial \Omega_3$ surround an area $\Omega$, where Ineqs.~\eqref{eq-4a}--\eqref{eq-4c} are all satisfied and stable limit cycles emerge:
\begin{eqnarray}
\Omega := \left\{ (r,a) \in \mathbb{R} \times \mathbb{R}_{>0} \mid a>a_{\mathrm{inf}},\ r_{\mathrm{inf}} < r < r_{\mathrm{sup}} \right\}.\label{Omega}
\end{eqnarray}\label{eq-6}
In $\Omega$, four qualitatively different types of oscillation patterns appear, as shown in Fig.~\ref{fig-t-xyz}(a)--(d);
\begin{itemize}
\setlength{\leftskip}{10pt}
\item Oscillation type (a) shown in Fig.~\ref{fig-t-xyz}(a)\\
Of those discussed in previous studies, the most typical is this, where the neurons alternate between the firing and resting modes.
\item Oscillation type (b) shown in Fig.~\ref{fig-t-xyz}(b)\\
This is a relatively small and fast oscillation in the vicinity of $\partial \Omega_{1}$, where the oscillation period is virtually independent from $r$ and decreases as $a$ decreases.
\item Oscillation type (c) shown in Fig.~\ref{fig-t-xyz}(c)\\
This emerges near $\partial \Omega_{2}^+$ and $\partial \Omega_{3}^+$ when $a$ is greater than or equal to $a_{+}^*$. In the neighborhoods of these borderlines, a neurons firing duration is greatly extended. On the borderline $\partial \Omega_{2}^+$ or $\partial \Omega_{3}^+$, this neuron fires constantly while the other neuron remains quiescent, as shown in Fig.~\ref{fig-t-xyz}(f).
\item Oscillation type (d) shown in Fig.~\ref{fig-t-xyz}(d)\\
This is the remaining non-trivial oscillation type observed when $a < a_{+}^*$ and $r$ are near borderlines $\partial \Omega_{2}^-$ and $\partial \Omega_{3}^-$. This pattern has an invariant oscillation period under a fixed value of $a$. In this pattern, only one neuron is permanently excited while the other alternates between the firing and quiescent states.
\end{itemize}
As discussed later, there are additional borderlines $\Delta_1$ and $\Delta_2$ between the oscillation pattern (d) and the other patterns (a)--(c) on the $(r,a)$ space (see Fig.~\ref{fig-phase}). The borderline threshold value $a_{+}^*$ above is derived in Subsection~\ref{subsec5B}. Note that the three critical borderlines $\partial \Omega_{1}$, $\partial \Omega_{2}$ and $\partial \Omega_{3}$ defined by Eqs.~\eqref{eq-6a}--\eqref{eq-6c} can be also generalized to the asymmetric connection case $a_{12} \neq a_{21}$, where the values of $a_{\mathrm{inf}}$, $r_{\mathrm{inf}}$, and $r_{\mathrm{sup}}$ are in the asymmetric version given by Eqs.~\eqref{eq-a_inf}--\eqref{eq-r_sup}, respectively (displayed on the inset of Fig.~\ref{fig-p}).

\subsection{Approximation of Oscillation Period \label{subsec3B}}

\begin{figure*}
 \centering
 \includegraphics[width=1\linewidth]{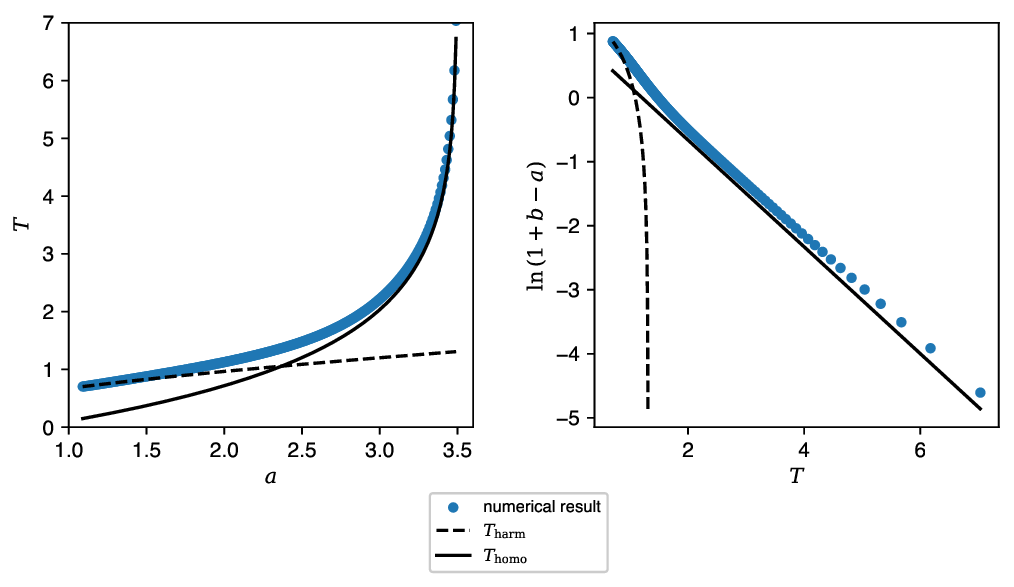}
 \caption{(Left) Plot of the oscillation period $T$ vs the symmetric synaptic weight $a$ with a fixed value $r=1$ corresponding to the dotted line (S1) in Fig.~\eqref{fig-p}. (Right) Another plot version of $\ln (1+b-a)$ vs the oscillation period $T$.}
 \label{fig-a-period}
\end{figure*}

Ref.~\onlinecite{matsuoka2011analysis} provided an approximated angular frequency of the limit cycle oscillation $\omega$ in the symmetric case $a=a_{12}=a_{21}$ and $r=\frac{s_2}{s_1}=1$:
\begin{eqnarray}
\omega = \frac{1}{\tau_y} \sqrt{\frac{(\tau_x+\tau_y)b-\tau_x a}{\tau_x a}}. \label{eq-matsuoka2011-omega}
\end{eqnarray}
Using this expression, we obtain the approximated oscillation period, $T_{\mathrm{harm}}$, as
\begin{eqnarray}
T_{\mathrm{harm}} := \frac{2 \pi}{\omega} = 2 \pi \tau_y \sqrt{\frac{\tau_x a}{(\tau_x+\tau_y)b-\tau_x a}}. \label{eq-matsuoka2011}
\end{eqnarray}
Eq.~\eqref{eq-matsuoka2011-omega} is based on the approximations of $x_i$ by a pure sinusoidal wave and the system \eqref{eq-3} by a harmonic oscillator. This is a good approximation when the nonlinear transformation from $x_i$ to $z_i$ given by Eq.~\eqref{z_i} does not change the overall waveform (i.e., $x_i \geq 0$ at almost all timepoints). This corresponds to a critical behavior as the oscillation type (b) seen in the vicinity of $\partial \Omega_1$. Figure.~\ref{fig-a-period} plots the actual oscillation periods (circle) and the approximated oscillation periods given by Eq.~\eqref{eq-matsuoka2011} (dotted line) along the cross-section (S1) are represented by the vertical dotted line in Fig.~\ref{fig-p}. As the value of $a$ decreases to $a_{\mathrm{inf}}$, the theoretical prediction $T_{\mathrm{harm}}$ approaches the numerically observed oscillation period $T$.

Two main problems arise concerning the approximation form $T_{\mathrm{harm}}$ given by Eq.~\eqref{eq-matsuoka2011}. First, $T_{\mathrm{harm}}$ is less accurate when $a$ is sufficiently large to approach borderlines $\partial \Omega_2$ and $\partial \Omega_3$. This is because the larger the waveform distortion from $x_i$ to $z_i$, the poorer the prediction accuracy of Eq.~\eqref{eq-matsuoka2011-omega}, as mentioned in Ref.~\onlinecite{matsuoka2011analysis}. Second, $T_{\mathrm{harm}}$ can be only applied to the case $a=a_{12}=a_{21}$ and $r=1$, which corresponds to the line (S1) in Fig.~\ref{fig-p}. This means that asymmetric oscillations like the oscillation type (c) cannot be well evaluated by the approximation form $T_{\mathrm{harm}}$.

Numerical observations imply the logarithmic divergence of the oscillation period under the larger value of $a$. In Subsection~\ref{subsec5C}, we propose a novel scaling law given by $T_{\mathrm{homo}}$, which can roughly approximate this logarithmic divergence [as represented by solid lines in Fig.~\ref{fig-a-period}]. We see that this approximation $T_{\mathrm{homo}}$ is also applicable to the cases of $a_{12} \neq a_{21}$ and $r \neq 1$, unlike the previous approximation $T_{\mathrm{harm}}$.

\section{EXISTENCE AND STABILITY OF FIXED POINTS \label{sec4}}

Within a single linear dynamical system, only a neutrally stable oscillation around a center is possible and no stable limit cycle emerges as a periodic solution. Nevertheless, a typical class of hybrid systems, the piecewise affine system, which consists of several discretely combined linear systems, sometimes possesses nontrivial solutions of stable limit cycles. The Matsuoka oscillator model is an example of the piecewise affine systems, as shown in Section~\ref{sec4}, because the system dynamics switch discontinuously across the $x_1$-axis and $x_2$-axis that represent the boundaries $x_i=0$ of Eq.~\eqref{z_i}.

Now, we apply fixed point analyses to the $n=2$ model \eqref{eq-3} for a precise discussion regarding the existence and stability of fixed points, which corresponds to stationary neuronal activities observed in Fig.~\ref{fig-t-xyz}(e), (f). As we see later, these analyses are essential for understanding the generation and bifurcation mechanism of limit cycle oscillations and deriving the oscillation period prediction \eqref{eq-approx}. For convenience, we represent a system state or a phase point $\bm{X}$ in the phase space $\mathbb{R}^4$:
\begin{eqnarray}
\bm{X} &:=& \begin{bmatrix}x_1 & y_1 & x_2 & y_2\end{bmatrix}^\top. \label{eq-X}
\end{eqnarray}
We also define an operator $[ \cdot ]_{q}$ that extracts the $q$-coordinate of the variable (e.g. $[ \bm{X} ]_{y_1}=y_1$).

Depending on the positivity or non-positivity of $x_1$ and $x_2$, we can divide the phase space $\mathbb{R}^4$ into four different regions, which correspond to four orthants on the $x_1$-$x_2$ plane
\begin{eqnarray}
\left.
\begin{aligned}
S_A &:=& \left\{ \bm{X} \in \mathbb{R}^4 \mid x_1, x_2 \leq 0 \right\}, \\
S_B &:=& \left\{ \bm{X} \in \mathbb{R}^4 \mid x_1>0,\ x_2 \leq 0  \right\}, \\
S_C &:=& \left\{ \bm{X} \in \mathbb{R}^4 \mid x_1 \leq 0,\ x_2 > 0 \right\}, \\
S_D &:=& \left\{ \bm{X} \in \mathbb{R}^4 \mid x_1, x_2 > 0 \right\}.
\end{aligned}
\right.
\end{eqnarray}
These regions are divided by the discontinuity boundaries below, which are defined as three-dimensional manifolds corresponding to parts of the $x_1$- and $x_2$-axes
\begin{eqnarray}
\left.
\begin{aligned}
\Sigma_{AB} = \Sigma_{BA} &:=& \left\{ \bm{X} \in \mathbb{R}^4 \mid x_1=0,\ x_2 \leq 0 \right\}, \\
\Sigma_{BD} = \Sigma_{DB} &:=& \left\{ \bm{X} \in \mathbb{R}^4 \mid x_1>0,\ x_2 = 0 \right\}, \\
\Sigma_{AC} = \Sigma_{CA} &:=& \left\{ \bm{X} \in \mathbb{R}^4 \mid x_1\leq0,\ x_2 = 0  \right\}, \\
\Sigma_{CD} = \Sigma_{DC} &:=& \left\{ \bm{X} \in \mathbb{R}^4 \mid x_1=0,\ x_2 > 0 \right\}, \\
\end{aligned}
\right.
\end{eqnarray}
Note that through this paper, unless there is any confusion, the discontinuity boundaries in the phase space above are all termed as ``boundaries,'' which are distinguished from ``borderlines'' such as $\partial \Omega_1$ and $\Delta_1$ in the $(r,a)$ parameter space. Although notation couples such as $\Sigma_{CD}$ and $\Sigma_{DC}$ refer to the same switching manifold, in this study, they are labeled separately depending on which direction the system state follows through this manifold. For example, when the system state transits from $S_C$ into $S_D$, it is designated as crossing $\Sigma_{CD}$; when it goes from $S_D$ into $S_C$, it is represented as crossing $\Sigma_{DC}$.

Similar to the single neuron analysis (Subsection~\ref{subsec2A}), the two-neuron system \eqref{eq-3} can be represented as a set of four linear systems by replacing $z_i$ with either $x_i$ or $0$ according to Eq.~\eqref{z_i}. The linear dynamics of each is defined within one of the divided regions. For instance, within region $S_i$ ($i \in \{ A, B, C, D \}$), the dynamic is
\begin{eqnarray}
\frac{d\bm{X}}{dt}=J_i\bm{X}+\bm{s}, \label{eq-dynamics}
\end{eqnarray}
where $\bm{s}=\begin{bmatrix} \frac{s_1}{\tau_x} & 0 & \frac{s_2}{\tau_x} & 0 \end{bmatrix}^\top$. A fixed point $\bm{X}_i^*$ of these dynamics meet $\frac{d\bm{X}}{dt}=\bm{0}$, or
\begin{eqnarray}
J_i\bm{X}_i^*+\bm{s}=\bm{0}, \label{eq-fixedpoint}
\end{eqnarray}
the stability of which is characterized by the eigenvalues of matrix $J_i$; if at least one of the eigenvalues $\lambda_i$ has a positive real part, the fixed point $\bm{X}_i^*$ is unstable because the trajectories of the system repelled from it in the corresponding eigendirection. Conversely, when the real parts of all eigenvalues are negative, the fixed point is stable.\cite{strogartz1994nonlinear} In this sense, $J_i$ is equivalent to the Jacobian matrix around the fixed point $\bm{X}_i^*$. Note that the stability here means linear and asymptotic stability.

\begin{figure}
 \includegraphics[width=\linewidth]{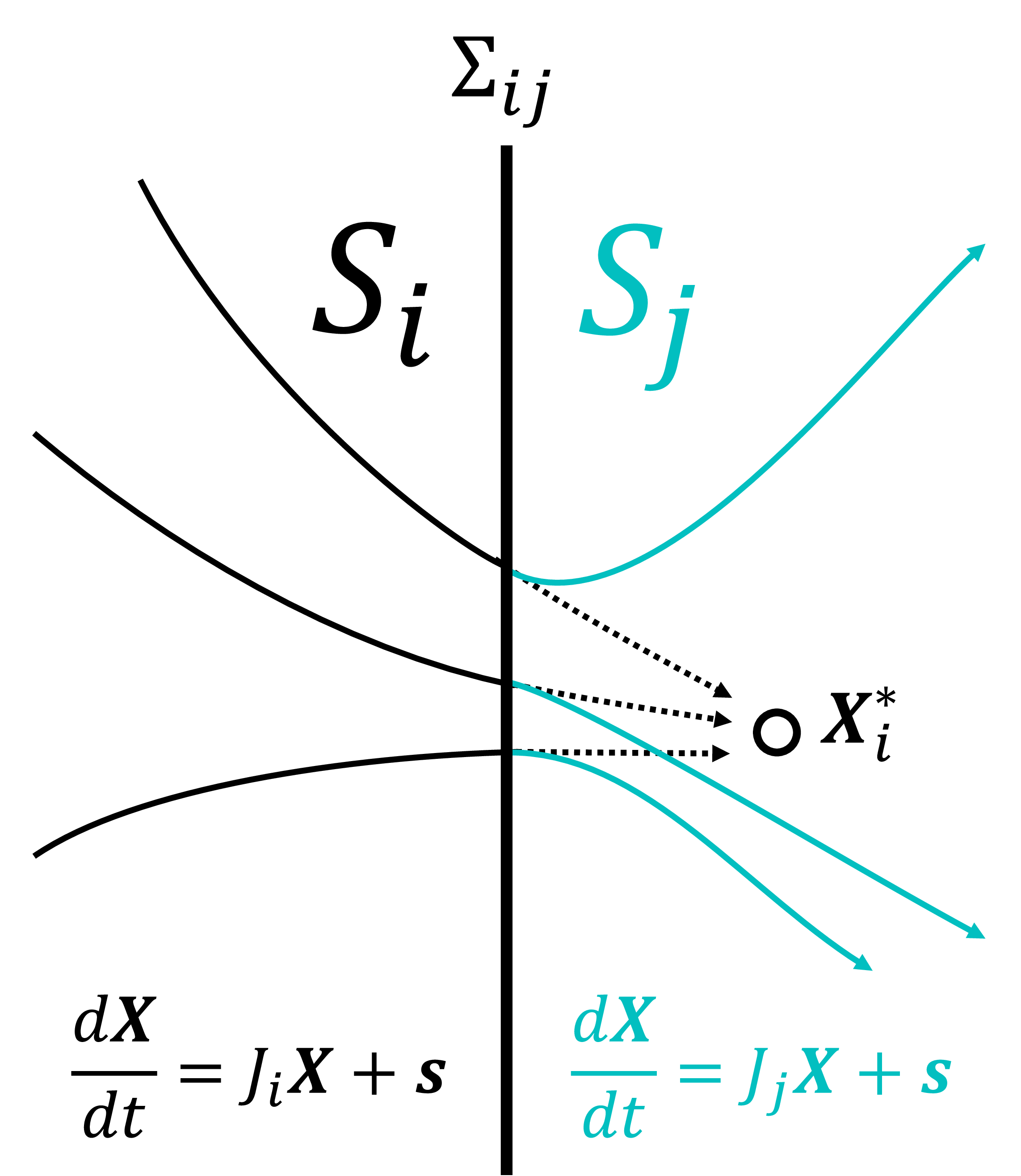}
 \caption{Conceptual example of a virtual fixed point. Suppose that a fixed point $\bm{X}_i^*$ determined with respect to dynamics $\frac{d\bm{X}}{dt}=J_i \bm{X} + \bm{s}$ is stable but virtual because $\bm{X}_i^* \in S_j$; when the system state is in the region $S_i$, it is about to converge to $\bm{X}_i^*$; after crossing the boundary $\Sigma_{ij}$, however, the system state comes to follow the other dynamics $\frac{d\bm{X}}{dt}=J_j \bm{X} + \bm{s}$; thus, $\bm{X}_i^*$ no longer functions as a stable fixed point in $S_j$.}
 \label{fig-virtual}
\end{figure}

A key point is that the fixed point $\bm{X}_i^*$ is determined corresponding to the dynamics $\frac{d\bm{X}}{dt}=J_i\bm{X}+\bm{s}$, independently of the region $S_i$. This means that $\bm{X}_i^* \in S_i$ is not always true. Suppose that
\begin{eqnarray}
\bm{X}_i^* \in S_j \ (j \neq i), \label{eq-virtual}
\end{eqnarray}
then $\bm{X}_i^*$ does not exist, or in other words, is a ``virtual'' fixed point\cite{di2008bifurcations} because the system state at $\bm{X}_i^*$ can follow
\begin{eqnarray}
\frac{d\bm{X}}{dt}=J_j\bm{X}_i^* +\bm{s} \neq \bm{0},
\end{eqnarray}
so that the system is no longer stationary in this situation (see also Fig.~\ref{fig-virtual}). Conversely, if
\begin{eqnarray}
\bm{X}_i^* \in S_i, \label{eq-regular}
\end{eqnarray}
then the fixed point $\bm{X}_i^*$ is regarded as ``regular'' (or ``admissible'') because it exists.\cite{di2008bifurcations, bernardo2008piecewise} Differently expressed, the condition \eqref{eq-regular} provides the existence condition of $\bm{X}_i^*$. In the following subsections, the four cases of dynamics are investigated for each index $i \in \{ A, B, C, D \}$, regarding
\begin{itemize}
\item the Jacobian matrix $J_i$,
\item the fixed point $\bm{X}_i^*$,
\item the eigenvalues $\lambda_i$, and
\item the existence condition $\bm{X}_i^* \in S_i.$
\end{itemize}

\subsection{Dynamics in the Region $S_A$ \label{subsec4A}}

\begin{subequations}
\begin{eqnarray}
J_A &=&
\begin{bmatrix}
-\frac{1}{\tau_x} & -\frac{b}{\tau_x} & 0 & 0 \\
&&&\\
0 & -\frac{1}{\tau_y} & 0 & 0 \\
&&&\\
0 & 0 & -\frac{1}{\tau_x} & -\frac{b}{\tau_x} \\
&&&\\
0 & 0 & 0 & -\frac{1}{\tau_y}
\end{bmatrix}, \label{eq-Aa}\\
\bm{X}_A^* &=&
\begin{bmatrix}
s_1 & 0 & rs_1 & 0 
\end{bmatrix}^\top, \label{eq-Ab}\\
\lambda_A &=& -\frac{1}{\tau_x}, -\frac{1}{\tau_y}. \label{eq-Ac}
\end{eqnarray}

This case is the simplest because neither neuron fires, resulting in no reciprocal interaction between the two. In this sense, each neuron can be treated as an independent single neuron. The existence condition $\bm{X}_A^* \in S_A$ is
\begin{eqnarray}
s_1, rs_1 \leq 0.\label{eq-Ad}
\end{eqnarray}
It is obvious that $\bm{X}_A^*$ is stable because the eigenvalues \eqref{eq-Ac} are negative real numbers.
\label{eq-A}
\end{subequations}

\subsection{Dynamics in the Region $S_B$ \label{subsec4B}}

\begin{subequations}
\begin{eqnarray}
J_B &=&
\begin{bmatrix}
-\frac{1}{\tau_x} & -\frac{b}{\tau_x} & 0 & 0 \\
&&&\\
\frac{1}{\tau_y} & -\frac{1}{\tau_y} & 0 & 0 \\
&&&\\
-\frac{a_{21}}{\tau_x} & 0 & -\frac{1}{\tau_x} & -\frac{b}{\tau_x} \\
&&&\\
0 & 0 & 0 & -\frac{1}{\tau_y}
\end{bmatrix}, \label{eq-Ba}\\
\bm{X}_B^* &=&
\begin{bmatrix}
\frac{s_1}{1+b} & \frac{s_1}{1+b} & \frac{(1+b)r-a_{21}}{1+b}s_1 & 0 
\end{bmatrix}^\top, \label{eq-Bb}\\
\lambda_B &=& -\frac{1}{\tau_x}, -\frac{1}{\tau_y}, \frac{-\tau_x-\tau_y \pm \sqrt{Q_B}}{2\tau_x\tau_y}, \label{eq-Bc}
\end{eqnarray}
where $Q_B = (\tau_y-\tau_x)^2-4b\tau_x\tau_y$. The existence condition $\bm{X}_B^* \in S_B$ is
\begin{eqnarray}
s_1>0 \quad \mathrm{and} \quad r \leq \frac{a_{21}}{1+b}. \label{eq-Bd}
\end{eqnarray}
\label{eq-B}
\end{subequations}

In this case, neuron-2 is significantly more strongly inhibited by neuron-1 than the direct external input $s_2$; therefore, with only the neuron-1 activated, the neuron-2 no longer fires. Two of the eigenvalues $\lambda_B = -\frac{1}{\tau_x}, -\frac{1}{\tau_y}$ are both negative real numbers, and the same holds for the remaining two $\lambda_B = \frac{-\tau_x-\tau_y \pm \sqrt{Q_B}}{2\tau_x\tau_y}$ because of the non-damped oscillation condition \eqref{eq-2}. Therefore, $\bm{X}_B^*$ is a stable node. The neuronal activity pattern shown in Fig.~\ref{fig-t-xyz}(f) corresponds to the convergence to $\bm{X}_B^*$.

\subsection{Dynamics in the Region $S_C$ \label{subsec4C}}

\begin{subequations}
\begin{eqnarray}
J_C &=&
\begin{bmatrix}
-\frac{1}{\tau_x} & -\frac{b}{\tau_x} & -\frac{a_{12}}{\tau_x} & 0 \\
&&&\\
0 & -\frac{1}{\tau_y} & 0 & 0 \\
&&&\\
0 & 0 & -\frac{1}{\tau_x} & -\frac{b}{\tau_x} \\
&&&\\
0 & 0 & \frac{1}{\tau_y} & -\frac{1}{\tau_y}
\end{bmatrix}, \label{eq-Ca}\\
\bm{X}_C^* &=&
\begin{bmatrix}
\frac{(1+b)-ra_{12}}{1+b}s_1 & 0 & \frac{rs_1}{1+b} & \frac{rs_1}{1+b} 
\end{bmatrix}^\top, \label{eq-Cb}\\
\lambda_C &=& -\frac{1}{\tau_x}, -\frac{1}{\tau_y}, \frac{-\tau_x-\tau_y \pm \sqrt{Q_C}}{2\tau_x\tau_y}, \label{eq-Cc}
\end{eqnarray}
where $Q_C = (\tau_y-\tau_x)^2-4b\tau_x\tau_y$. This case is completely symmetric to the dynamics in the region $S_B$, with indices 1 and 2 switching to each other. The existence condition of $\bm{X}_C^* \in S_C$ is
\begin{eqnarray}
s_2=rs_1 > 0 \quad \mathrm{and} \quad \frac{1}{r} \leq \frac{a_{12}}{1+b}. \label{eq-Cd}
\end{eqnarray}
\label{eq-C}
\end{subequations}

\subsection{Dynamics in the Region $S_D$ \label{subsec4D}}

\begin{subequations}
\begin{eqnarray}
J_D &=&
\begin{bmatrix}
-\frac{1}{\tau_x} & -\frac{b}{\tau_x} & -\frac{a_{12}}{\tau_x} & 0 \\
&&&\\
\frac{1}{\tau_y} & -\frac{1}{\tau_y} & 0 & 0 \\
&&&\\
-\frac{a_{21}}{\tau_x} & 0 & -\frac{1}{\tau_x} & -\frac{b}{\tau_x} \\
&&&\\
0 & 0 & \frac{1}{\tau_y} & -\frac{1}{\tau_y}
\end{bmatrix}. \label{eq-Da}
\end{eqnarray}

This is the most complicated case and requires a fixed-point analysis of $\bm{X}_D^*$. If $(1+b)^2 \neq a_{12}a_{21}$, then $\bm{X}_D^*$ is written as
\begin{eqnarray}
\bm{X}_D^* =
\frac{s_1}{(1+b)^2 - a_{12}a_{21}}
\begin{bmatrix}
(1+b) - ra_{12} \\
(1+b) - ra_{12} \\
(1+b)r - a_{21} \\
(1+b)r - a_{21}
\end{bmatrix}. \label{eq-Db}
\end{eqnarray}
In the remaining singular case $1+b = \sqrt{a_{12}a_{21}}$, the fixed point cannot be simply written as Eq.~\eqref{eq-Db}. If $(ra_{12} - \sqrt{a_{12}a_{21}})s_1 = 0$ in addition to $1+b = \sqrt{a_{12}a_{21}}$, a set of non-isolated fixed points is represented as a line in the four-dimensional system space:
\begin{equation}
\bm{X}_D^* = \left\{ \bm{X} \in \mathbb{R}^4 \ \middle| \ x_2=-\sqrt{\frac{a_{21}}{a_{12}}}x_1+\frac{1}{a_{12}}s_1, y_1=x_1, y_2=x_2 \right\}. \label{eq-Dc}
\end{equation}
Note that $\bm{X}_B^*$ and $\bm{X}_C^*$ are both on the line \eqref{eq-Dc} in these conditions. By contrast, no solution satisfies $\frac{d\bm{X}}{dt}=0$ when $(ra_{12} - \sqrt{a_{12}a_{21}})s_1 \neq 0$.

Whether or not the fixed point $\bm{X}_D^*$ is written as Eq.~\eqref{eq-Db} or \eqref{eq-Dc}, its stability is determined by the eigenvalues $\lambda_D$, which are solutions of the characteristic equation $\det(\lambda_D I-J_D)=0$. Here,
\begin{eqnarray}
\begin{split} 
& \det(\lambda_D I-J_D) \\
& = \left(\lambda_D^2+\frac{\tau_x+\tau_y+\tau_y\sqrt{a_{12}a_{21}}}{\tau_x\tau_y}\lambda_D+\frac{1+b+\sqrt{a_{12}a_{21}}}{\tau_x\tau_y} \right) \\
& \times \left(\lambda_D^2+\frac{\tau_x+\tau_y-\tau_y\sqrt{a_{12}a_{21}}}{\tau_x\tau_y}\lambda_D+\frac{1+b-\sqrt{a_{12}a_{21}}}{\tau_x\tau_y} \right).
\end{split} \label{eq-lambdaD}
\end{eqnarray}
The real parts of the first two eigenvalues represented by $\lambda_{D+}$ as the solutions of
\begin{eqnarray}
\lambda_{D+}^2+\frac{\tau_x+\tau_y+\tau_y\sqrt{a_{12}a_{21}}}{\tau_x\tau_y}\lambda_{D+}+\frac{1+b+\sqrt{a_{12}a_{21}}}{\tau_x\tau_y} =0, \label{eq-D+}
\end{eqnarray}
are always negative, which suggests that the dynamics converge to $\bm{X}_D^*$ on the plane spanned by the two corresponding eigenvectors $\bm{v}_{D+}$.

The remaining two eigenvalues $\lambda_{D-}$ are given by the equation
\begin{eqnarray}
\lambda_{D-}^2+\frac{\tau_x+\tau_y-\tau_y\sqrt{a_{12}a_{21}}}{\tau_x\tau_y}\lambda_{D-}+\frac{1+b-\sqrt{a_{12}a_{21}}}{\tau_x\tau_y}=0. \label{eq-D-}
\end{eqnarray}
The two corresponding eigenvectors $\bm{v}_{D-}$ span a plane on the phase space
\begin{equation}
P_D := \left\{ \bm{X} \in \mathbb{R}^4 \ \middle| \ x_2=-\sqrt{\frac{a_{21}}{a_{12}}}x_1+\frac{r+\sqrt{\frac{a_{21}}{a_{12}}}}{1+b+\sqrt{a_{12}a_{21}}}s_1, y_2=-\sqrt{\frac{a_{21}}{a_{12}}}y_1+\frac{r+\sqrt{\frac{a_{21}}{a_{12}}}}{1+b+\sqrt{a_{12}a_{21}}}s_1 \right\}. \label{eq-Dc}
\end{equation}
According to the discussion above, the system state outside $P_D$ is asymptotically attracted to $P_D$ through the stable eigendirections $\bm{v}_{D+}$. This means that the stability of $\bm{X}_D^*$ is finally determined as the stability on the two-dimensional plane $P_D$, according to the values of $\bm{\lambda}_{D-}$;\cite{strogartz1994nonlinear}
\begin{itemize}
\setlength{\leftskip}{10pt}
\item $\bm{X}_D^*$ is a stable node or spiral if
\begin{eqnarray}
\sqrt{a_{12}a_{21}} < \min\left(1+\frac{\tau_x}{\tau_y}, 1+b\right). \label{eq-D1}
\end{eqnarray}
Figure.~\ref{fig-t-xyz}(e) is the neuronal activity converging to $\bm{X}_D^*$, where $\bm{X}_D^*$ is a regular stable spiral.
\item $\bm{X}_D^*$ becomes a center if
\begin{eqnarray}
\sqrt{a_{12}a_{21}} = 1+\frac{\tau_x}{\tau_y} < 1+b. \label{eq-D2}
\end{eqnarray}
Conservative oscillation occurs around it on the plane spanned by the two corresponding eigenvectors.
\item $\bm{X}_D^*$ behaves as an unstable node or spiral, if
\begin{eqnarray}
1+\frac{\tau_x}{\tau_y}<\sqrt{a_{12}a_{21}}<1+b, \label{eq-D3}
\end{eqnarray}
which repels from the remaining two eigendirections if
\item $\bm{X}_D^*$ comprises non-isolated fixed points written as Eq.~\eqref{eq-Dc}, if
\begin{eqnarray}
\sqrt{a_{12}a_{21}}=1+b, \label{eq-D4}
\end{eqnarray}
where one eigenvalue, $\lambda_{D-}$, is 0 and the corresponding eigenvector is parallel to the line \eqref{eq-Dc}. This is equivalent to the non-isolated fixed point as the line given by Eq.~\eqref{eq-Dc}. Additionally, in the remaining eigendirection, $\bm{X}_D^*$ attracts if $b < \frac{\tau_x}{\tau_y}$, and repels if $\frac{\tau_x}{\tau_y} < b$. If $\sqrt{a_{12}a_{21}}=1+b=1+\frac{\tau_x}{\tau_y}$, then $\bm{X}_D^*$ becomes further non-isolated fixed points as a plane equal to $P_D$.
\item $\bm{X}_D^*$ is a saddle point if
\begin{eqnarray}
1+b<\sqrt{a_{12}a_{21}}. \label{eq-D5}
\end{eqnarray}
Trajectories are repelled from $\bm{X}_D^*$ in a single eigendirection corresponding to a positive eigenvalue $\lambda_{D-}$.
\end{itemize}

The existence condition $\bm{X}_D^* \in S_D$ is also separately discussed depending on the relation between $1+b$ and $ \sqrt{a_{12}a_{21}}$.
\begin{itemize}
\setlength{\leftskip}{10pt }
\item When $1+b > \sqrt{a_{12}a_{21}}$, $\bm{X}_D^* \in S_D$ is true if and only if
\begin{eqnarray}
\frac{1+b}{a_{12}}>r>\frac{a_{21}}{1+b}>0 \quad \mathrm{and} \quad s_1>0. \label{eq-Dd}
\end{eqnarray}
In this situation, the necessary and sufficient condition below is also true.
\begin{eqnarray}
\bm{X}_A^* \notin A \ \mathrm{and} \ \bm{X}_B^* \notin B \ \mathrm{and} \ \bm{X}_C^* \notin C \ \Longleftrightarrow \ \bm{X}_D^* \in S_D. \label{eq-De}
\end{eqnarray}
\item When $1+b = \sqrt{a_{12}a_{21}}$, $\bm{X}_D^* \in S_D$ is true if and only if
\begin{eqnarray}
r=\frac{1+b}{a_{12}}=\frac{a_{21}}{1+b}>0 \quad \mathrm{and} \quad s_1>0. \label{eq-Df}
\end{eqnarray}
Note that the repelling eigendirection of $\bm{X}_D^*$ leads to $\bm{X}_B^*$ or $\bm{X}_C^*$.
\item When $1+b < \sqrt{a_{12}a_{21}}$, $\bm{X}_D^* \in S_D$ is true if and only if
\begin{eqnarray}
\frac{a_{21}}{1+b}>r>\frac{1+b}{a_{12}}>0 \quad \mathrm{and} \quad s_1>0. \label{eq-Dg}
\end{eqnarray}
This agrees with the condition under which $\bm{X}_D^*$ becomes a saddle point.
\end{itemize}
In all cases, a necessary condition for $\bm{X}_D^* \in S_D$ is described as
$$s_1,r  > 0.$$
Consequently, the external inputs into the neuron-1 and neuron-2 are both excitatory.
\end{subequations}

\section{BIFURCATIONS BETWEEN FIXED POINTS AND OSCILLATORY SOLUTIONS \label{sec5}}

\begin{figure}[b]
\includegraphics[width=\linewidth]{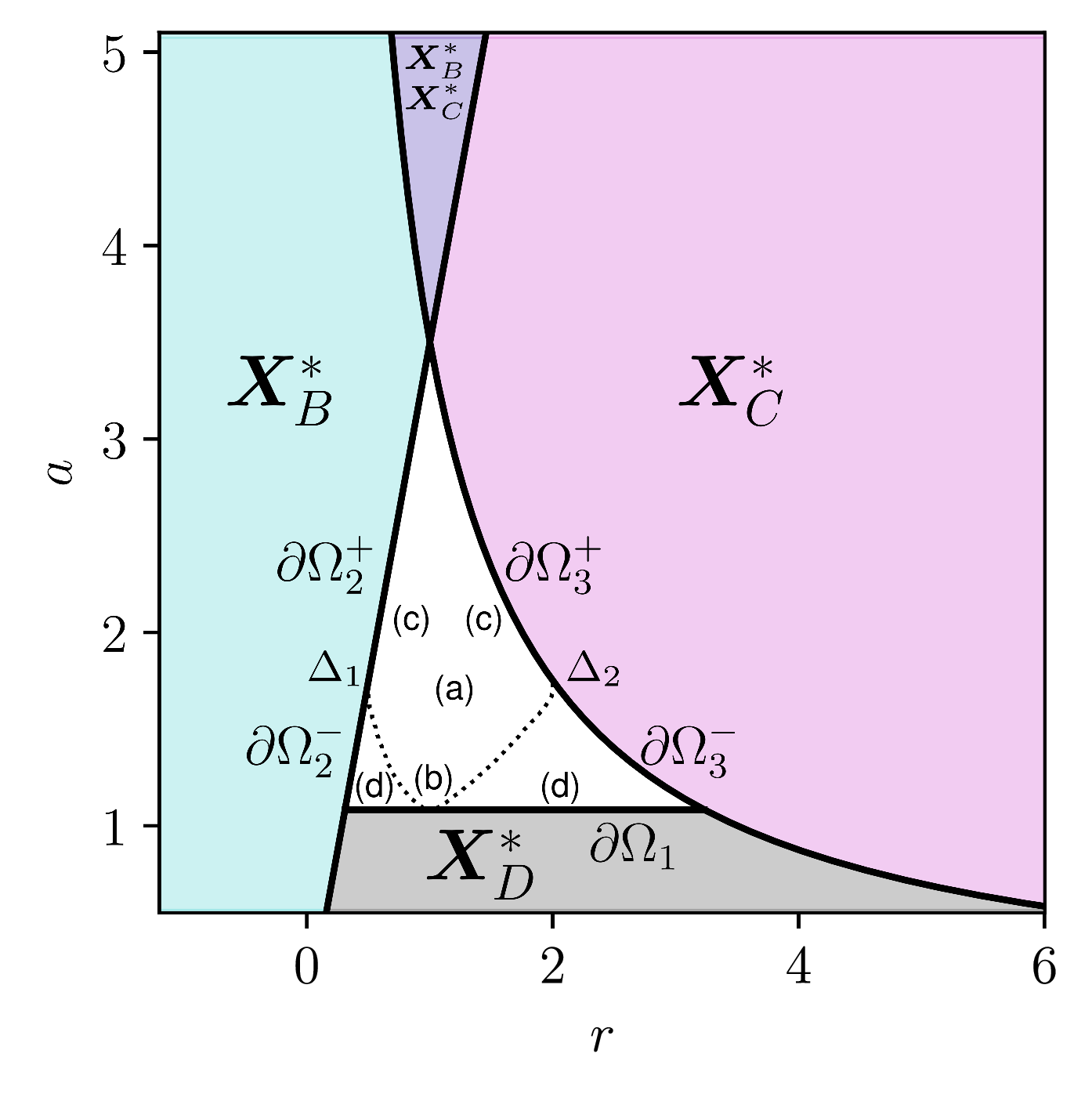}
\caption{Phase diagram of the same $r$-$a$ plane as in Fig.~\ref{fig-p}. Stable oscillations emerge in $\Omega$ (white) surrounded by the borderlines $\partial \Omega_1$, $\partial \Omega_2$, and $\partial \Omega_3$ (black solid lines). The notations (a)--(d) express the rough locations where the corresponding oscillation types (a)--(d) are observed. Note that the oscillation type (d) non-smoothly transits to the other oscillation types (a)--(c) at the $\Delta_1$ and $\Delta_2$ borderlines (dotted line).}
\label{fig-phase}
\end{figure}

According to Ineqs.~\eqref{eq-Ad}, the stable fixed point $\bm{X}_A^*$ does not exist when the two neurons are both suppressed by non-positive inputs. The other three fixed points $\bm{X}_B^*$, $\bm{X}_C^*$, $\bm{X}_D^*$ can exist when at least $s_1$ or $s_2=r s_1$ is positive. Specifically,
\begin{itemize}
\setlength{\leftskip}{10pt } 
\item the existence conditions $\bm{X}_B^* \in S_B$ given by \eqref{eq-Bd},
\item the existence conditions $\bm{X}_C^* \in S_C$ given by \eqref{eq-Bc}, and
\item the conditions \eqref{eq-D1} and \eqref{eq-Dd} under which $\bm{X}_D^* \in S_D$ is a stable fixed point,
\end{itemize}
are all separated from the existence condition of the oscillatory solutions Ineqs.~\eqref{eq-5a}--\eqref{eq-5c}. Otherwise expressed, if any oscillation occurs under the conditions \eqref{eq-5a}--\eqref{eq-5c}, then
\begin{eqnarray}
\bm{X}_B^*, \bm{X}_C^*, \bm{X}_D^* \in S_D, \label{eq-16}
\end{eqnarray}
which means that only $\bm{X}_D^*$ is regular while $\bm{X}_B^*$ and $\bm{X}_C^*$ are virtual fixed points. Satisfying Ineqs.~\eqref{eq-5a}--\eqref{eq-5c} leads to the necessary condition \eqref{eq-add1}, which is equivalent to Ineqs.~\eqref{eq-D3}. Hence, the existing fixed point $\bm{X}_D^*$ is an unstable node or spiral. Figure~\ref{fig-phase}, which shows the same $r$-$a$ plane as Fig.~\ref{fig-p}, is a phase diagram specifying the parameter regions where either $\bm{X}_B^*$, $\bm{X}_C^*$, $\bm{X}_D^*$, or oscillatory solutions of the types (a)--(d) exist as attractors. The overlapped purple area of Fig.~\ref{fig-phase} satisfies both \eqref{eq-Bd} and \eqref{eq-Cd}; thus, stable fixed points $\bm{X}_B^*$ and $\bm{X}_C^*$ exist simultaneously.

\subsection{Emergence Mechanism of Oscillatory Solutions \label{subsec5A}}

\begin{figure*}
\includegraphics[width=0.75\linewidth]{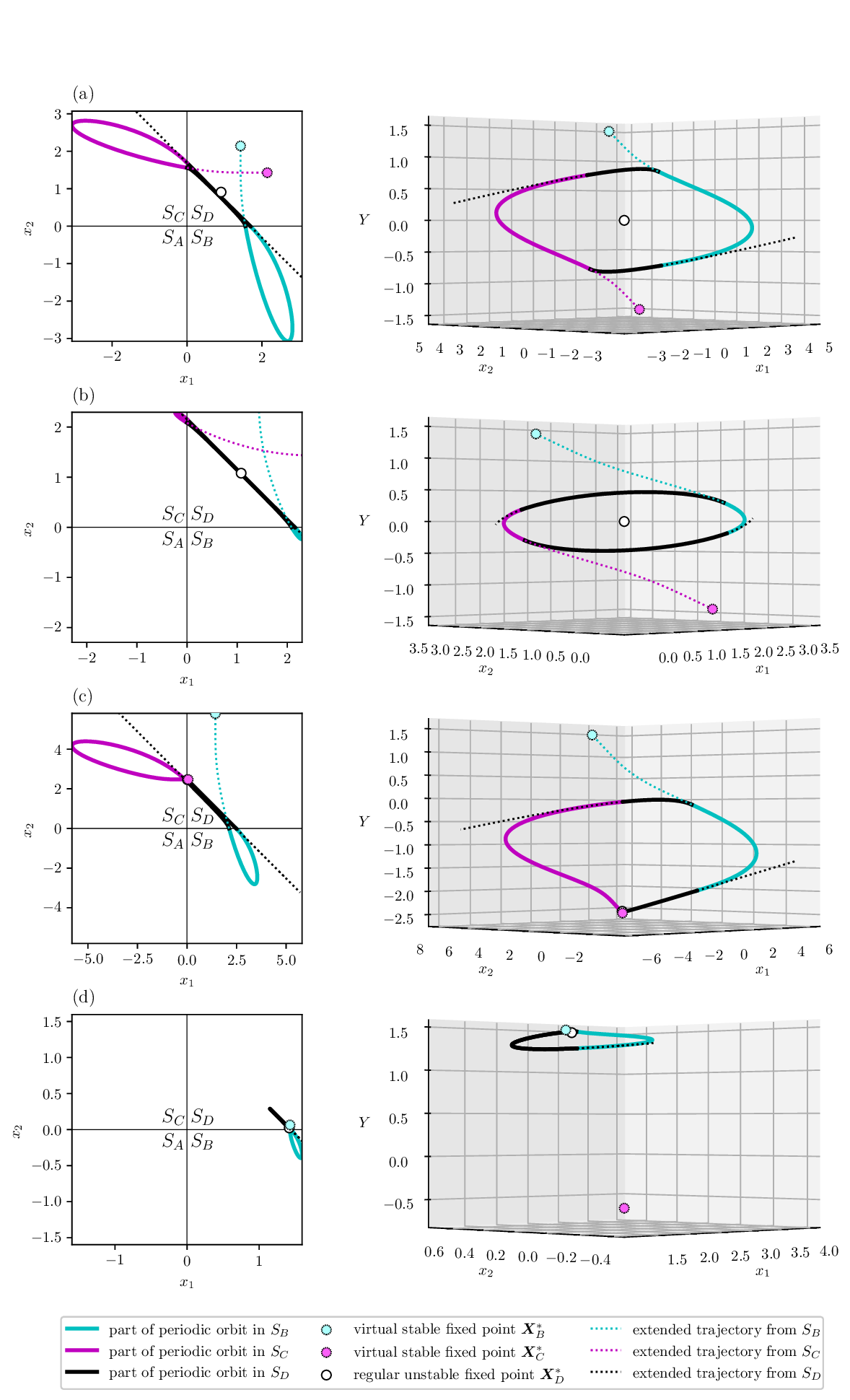}
\caption{Trajectories of oscillation patterns of Fig.~\ref{fig-t-xyz}(a)--(d) plotted on the $x_1$-$x_2$ plane (left column) and the $x_1$-$x_2$-$Y$ space (left column, for $Y=y_1-y_2$). The system dynamics switch at $\Sigma_{BD}$ (positive part of the $x_1$-axis) and $\Sigma_{CD}$ (positive part of the $x_2$-axis), where the color of the periodic orbit (solid line) changes. For $i \in \left\{B, C, D\right\}$, the ``extended trajectory from $S_i$'' (dotted line) would be hypothetically realized if the dynamics in the region $S_i$  continued to operate after the system state escaped from $S_i$.}
\label{fig-limitcycle}
\end{figure*}

Figure~\ref{fig-limitcycle} displays the same oscillation patterns (a)--(d) as in Fig.~\ref{fig-t-xyz}(a)--(d) on the $x_1$-$x_2$ plane (left column) and in the $x_1$-$x_2$-$Y$ space (right column) for $Y=y_1-y_2$. Note that the latter plot format is effective for observing overviews of the oscillation trajectories as previously proposed in Ref.~\onlinecite{matsuoka2011analysis}. Here, the emergence mechanism of oscillations under the conditions \eqref{eq-5a}--\eqref{eq-5c} is explained as repetitions of convergence to the stable but virtual $\bm{X}_B^*$, $\bm{X}_C^*$ and divergence from the regular but unstable $\bm{X}_D^*$. Since \eqref{eq-16}, if the system state is currently in the region $S_B$ (or $S_C$), it transits across the boundary $\Sigma_{BD}$ ($\Sigma_{CD}$) into the region $S_D$ before converging to $\bm{X}_B^*$ ($\bm{X}_C^*$) [the cyan (magenta) dotted lines shown in Fig.~\ref{fig-limitcycle} are the imaginary extended trajectories assumed if this convergence continues in the region $S_D$]. In the region $S_D$, the system state follows the dynamics $\frac{d\bm{X}}{dt}=J_D\bm{X}+\bm{s}$, in the eigendirection $\bm{v}_{D+}$ gets attracted to the plane $P_D$, and in turn, escapes from $S_D$ and crosses $\Sigma_{DC}$, $\Sigma_{DB}$, ($\Sigma_{DB}$, or $\Sigma_{DC}$). In the oscillation types (a)--(c), the system state transits across $\Sigma_{DC}$ ($\Sigma_{DB}$) into $S_C$ ($S_B$), after which the system state follows similar itineraries from $S_C$ ($S_B$) to $S_D$ and returns to $S_B$ ($S_C$):
$$S_B \to S_D \to S_C \to S_D \to S_B \to S_D \to S_C \to \cdots.$$
The remaining oscillation type (d) is exceptional because the system state immediately returns to $S_B$ ($S_C$) across $\Sigma_{DB}$ ($\Sigma_{DC}$); thus, the periodic orbit is captured in regions $S_D$ and $S_B$ ($S_C$) like
$$S_B \to S_D \to S_B \to S_D \to S_B \to S_D \to S_B \to \cdots,$$

Note that if the system state is currently in region $S_A$, it is immediately transferred into the regions $S_B$, $S_C$, or $S_D$ similar to the convergence to the virtual stable fixed point $\bm{X}_A^*$ outside $S_A$.

\subsection{Bifurcation Scenarios \label{subsec5B}}

\begin{figure*}
\includegraphics[width=\linewidth]{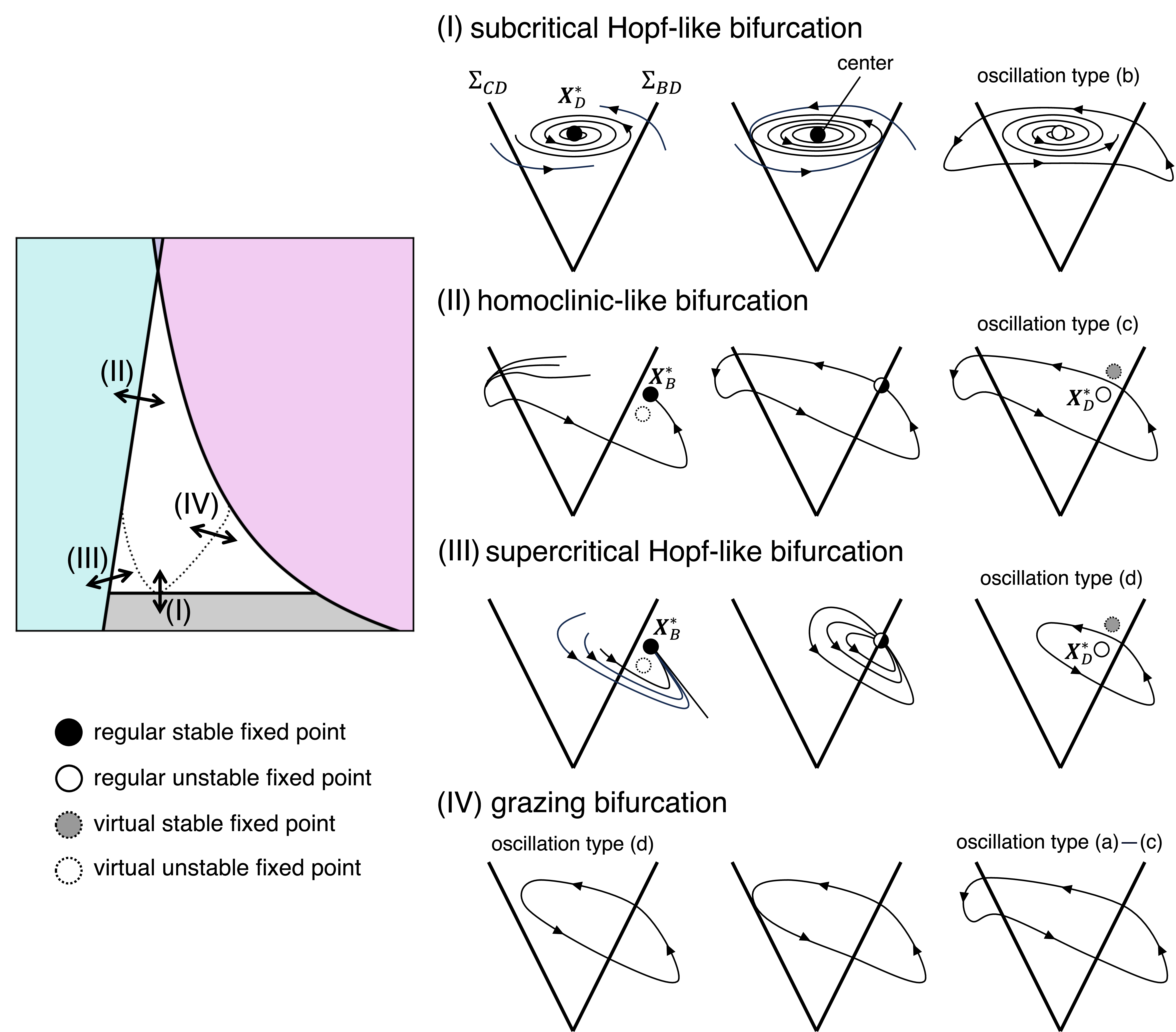}
\caption{Schematic diagram of the discontinuity-induced bifurcation scenarios. The bifurcations (I)--(IV) in the right-hand side are observed when parameter set $(r, a)$ changes crossing the borderlines $\partial \Omega_1$, $\partial \Omega_2$, or $\partial \Omega_1$, and $\Delta_1$ or $\Delta_2$ as shown in the left-hand side, which is the same phase diagram as Fig.~\ref{fig-phase}. In general, the periodic cycle generated through the subcritical Hopf-like bifurcation (I) is of both the oscillation types (b) and (d). In the bifurcation point of (I), the fixed point $X_D^*$ is a center. Although there are other regular and virtual fixed points in the bifurcations (I) and (IV), which are not directly related to the bifurcation scenarios, are omitted for simplicity.}
\label{fig-bifurcations}
\end{figure*}

The oscillation types (b)--(d) represent those observed in the neighborhoods of the borderlines $\partial \Omega_1$, $\partial \Omega_2$, and $\partial \Omega_3$ in the parameter space of Fig.~\ref{fig-phase}. Here, we can expect several bifurcation scenarios to emerge between these limit cycles and fixed points (Fig.~\ref{fig-bifurcations}). The first is a subcritical Hopf-like bifurcation observed on the borderline $\partial \Omega_1$, where the stability of $\bm{X}_D^*$ changes from an unstable spiral via a center to a stable spiral [Fig.~\ref{fig-bifurcations}(I)]. The previously proposed approximation of the oscillation period \eqref{eq-matsuoka2011} completely coincides in the actual oscillation period at the borderline $\partial \Omega_1$, which is identical to the harmonic oscillation period around the center $\bm{X}_D^*$. In this bifurcation point, the outermost periodic orbit grazing either or both $S_B$ and $S_C$ can be regarded as semihalf; it is neutrally stable from and toward its interior; and is attracting from its exterior because of the attracting flow in the outer regions $S_B$ and $S_C$ toward $S_D$. Beyond $\partial \Omega_1$ in the parameter space, the periodic orbits across multiple regions disappear because the system dynamics in the region $S_D$ are now incapable of carrying the system state through divergence outward to $S_B$ or $S_C$. Through this scenario, the limit cycle instantaneously emerges with non-zero (sufficiently large) value of oscillation amplitude, which is a canard-like behavior.

The second bifurcation is homoclinic-like, which emerges at $\partial \Omega_2^+$ and $\partial \Omega_3^+$. The oscillation type (c) is close to this bifurcation. In the vicinity of $\partial \Omega_2^+$ (or $\partial \Omega_3^+$), the virtual stable fixed point $\bm{X}_B^*$ ($\bm{X}_C^*$) becomes regular upon entering the region $S_B$ ($S_C$) on the pre-existing limit cycle orbit [Fig.~\ref{fig-bifurcations}(II)]. When the parameter set $(r,a)$ is on the $\partial \Omega_2^+$ ($\partial \Omega_3^+$), then $\bm{X}_B^*$ ($\bm{X}_C^*$) is precisely on the boundary $\Sigma_{BD}$ ($\Sigma_{CD}$) and identical to $\bm{X}_C^*$, which is termed a boundary equilibrium.\cite{di2008bifurcations, bernardo2008piecewise} Because $\bm{X}_B^*$ ($\bm{X}_C^*$) attracts from $S_B$ ($S_C$) and repels into $S_D$, it functions as a saddle-like point [this behavior is reflected in Fig.~\ref{fig-limitcycle}(c)]. At this bifurcation point, the regularity of $\bm{X}_D^*$ and the virtuality of $\bm{X}_B^*$ can be regarded as exchanged with each other. This is termed a persistence, a type of boundary-equilibrium bifurcation.\cite{di2008bifurcations, bernardo2008piecewise} At the intersection of $\partial \Omega_2$ and $\partial \Omega_3$, where $(r,a)=(1,1+b)$, this homoclinic-like bifurcation becomes codimension-two through which two homoclinic orbits are expected to merge into a heteroclinic orbit.

Finally, the bifurcation pattern corresponding to the oscillation type (d) emerges on $\partial \Omega_2^-$ and $\partial \Omega_3^-$. Similar to the homoclinic-like bifurcation, $\bm{X}_D^*$ disappears and $\bm{X}_B^*$ (or $\bm{X}_C^*$) comes to exist through a persistence [Fig.~\ref{fig-bifurcations}(III)], when $r$ or $a$ crosses the borderline $\partial \Omega_2^-$ ($\partial \Omega_3^-$). However, in this bifurcation, the limit cycle trajectories as the combination of divergence from $\bm{X}_D^*$ and convergence to $\bm{X}_B^*$ ($\bm{X}_C^*$) become extremely small [see also Fig.~\ref{fig-limitcycle}(d)], converging to zero oscillation amplitudes. Although the oscillation period is constant in the neighborhoods of the borderlines (mentioned in Section~\ref{subsec3A}), this bifurcation seems supercritical Hopf-like, or a so-called discontinuity-induced Hopf bifurcation.\cite{bernardo2008piecewise}

The second homoclinic-like bifurcation and the third supercritical Hopf-like bifurcation are both boundary-equilibrium bifurcations caused by regular-virtual exchanges between fixed points, through which their stability or instability is conserved. Corresponding to the (c) and (d) oscillation pattern, these bifurcations are distinguished by whether $\bm{X}_D^*$ works as an unstable node or an unstable spiral. If $\bm{X}_D^*$ is an unstable node, then the dynamics in the region $S_D$ certainly transport the system state from $\Sigma_{BD}$ (or $\Sigma_{CD}$) to the opposite side $\Sigma_{DC}$ (or $\Sigma_{DB}$), in the fastest eigendirection corresponding to one of $\lambda_{D-}$ with the largest absolute value [the lower black solid line shown in the right of Fig.~\ref{fig-limitcycle}(c) is almost straight, which is extremely close to this eigendirection]. Conversely, when $\bm{X}_D^*$ is an unstable spiral, the rotational dynamics around this point can afterward return the system state coming from $\Sigma_{BD}$ ($\Sigma_{CD}$) to the same side $\Sigma_{DB}$ ($\Sigma_{DC}$) [the black solid lines on the right-hand side of Figs.~\ref{fig-limitcycle}(b) and (d) are curling due to this spiral effect]. Therefore, the threshold $a=a_{+}^*$ between these two bifurcations is determined by the point at which $\bm{X}_D^*$ changes between a node and a spiral. This is where the discriminant $Q_{D-}$ of Eq.~\eqref{eq-D-} is 0. Solving an equation $Q_{D-}=0$ for $a$, we obtain two solutions as
\begin{eqnarray}
a=a_{+}^*, a_{-}^*, \label{eq-a_+-}
\end{eqnarray}
where
\begin{subequations}
\begin{eqnarray}
a_{+}^* := \frac{\tau_y-\tau_x + 2\sqrt{b\tau_x\tau_y}}{\tau_y}, \label{eq-a_+}\\
a_{-}^* := \frac{\tau_y-\tau_x - 2\sqrt{b\tau_x\tau_y}}{\tau_y}. \label{eq-a_-}
\end{eqnarray}
\end{subequations}
The former satisfies $a_{\mathrm{inf}}<a_{+}^*<1+b$, whereas the latter meets $0<a_{-}^*<a_{\mathrm{inf}}$. Thus, the threshold value $a_{+}^*$ holds by \eqref{eq-a_+}. The zenith of the dotted lines $\Delta_1$ and $\Delta_2$ on the vertical $a$-axis in Fig.~\ref{fig-phase} corresponds to $a=a_{+}^*$. Note that even if $\bm{X}_D^*$ is a spiral, however, the system state which entered $S_D$ crossing a boundary is not always destined to return to the same boundary. Suppose that the parameter condition is not in the neighborhood of $\partial \Omega_2^-$ and $\partial \Omega_3^-$, then $\bm{X}_D^*$ can be separated from the two boundaries $\Sigma_{BD}$ and $\Sigma_{DC}$. In this situation, it may be sufficient to bridge the system state from $\Sigma_{BD}$ ($\Sigma_{CD}$) to the opposite side $\Sigma_{DC}$ ($\Sigma_{DB}$), no matter how the spiral dynamics around $\bm{X}_D^*$ work. The borderline of whether the stable limit cycle grazes the boundary $\Sigma_{DC}$ ($\Sigma_{DB}$) and the system state thereby successfully bridges to the opposite side, is represented as the dotted lines $\Delta_1$ and $\Delta_2$ shown in Fig.~\ref{fig-phase}. At these borderlines, the oscillation type (d) and the others (a)--(c) are clearly separated, and the transition between them can be regarded as a grazing bifurcation\cite{di2008bifurcations} of the limit cycle [Fig.~\ref{fig-bifurcations}(IV)].

\subsection{Derivation of the Logarithmic Oscillation-Period Scaling Law \label{subsec5C}}

\begin{figure*}
 \centering
 \includegraphics[width=1\linewidth]{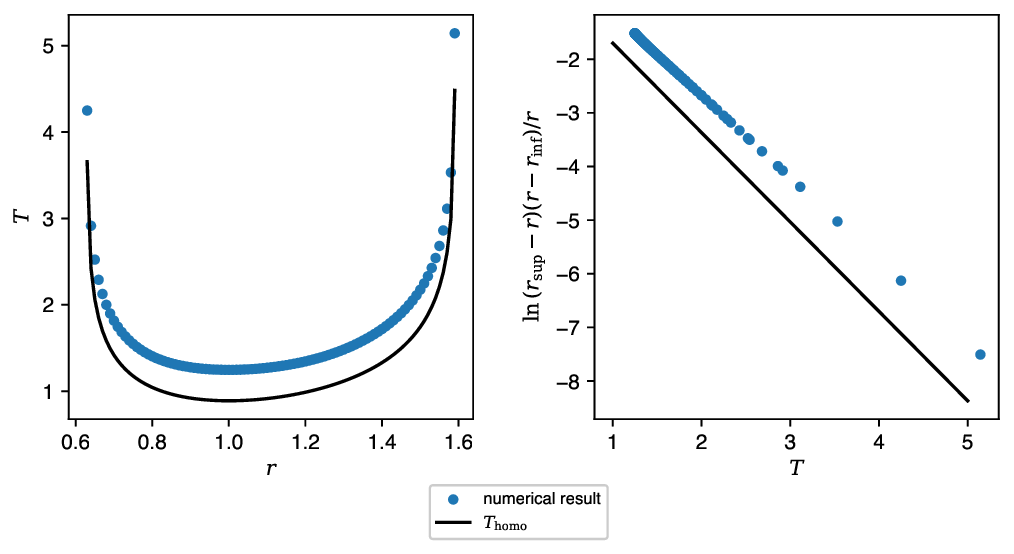}
 \caption{(Left) Plot of the oscillation period $T$ vs the symmetric synaptic weight $a$ with fixed value $a=2.2$ corresponding to the dotted line (S2) in Fig.~\eqref{fig-p}. (Right) Another plot version of $\ln (r_{\mathrm{sup}}-r)(r-r_{\mathrm{inf}})/r$ vs the oscillation period $T$.}
 \label{fig-a-r-p}
\end{figure*}

In Subsection~\ref{subsec3B}, we discuss the limitations of the approximated oscillation period $T_{\mathrm{harm}}$ given by Eq.~\eqref{eq-matsuoka2011}, which is previously proposed by Ref.~\onlinecite{matsuoka2011analysis}. Instead of $T_{\mathrm{harm}}$, we introduce a logarithmic scaling law $T_{\mathrm{homo}}$ as a more plausible approximation of oscillation period for larger asymmetric synaptic weights $a_{21} \neq a_{21}$ and asymmetric inputs given by $r \neq 1$:
\begin{eqnarray}
T_{\mathrm{homo}} := \tau_y \left[ \ln \frac{1}{(1+b)a_{12} \delta^2} - \ln \frac{(r_{\mathrm{sup}}-r)(r-r_{\mathrm{inf}})}{r} \right], \label{eq-approx}
\end{eqnarray}
where
\begin{eqnarray}
\delta := \frac{\tau_y-\tau_x}{b\tau_y}. \label{eq-delta}
\end{eqnarray}
These logarithmic forms of oscillation periods are commonly observed in the vicinity of homoclinic bifurcations.\cite{gaspard1990measurement, santos2004effects} Figure~\ref{fig-a-r-p} plots both the theoretical prediction $T_{\mathrm{homo}}$ by Eq.~\eqref{eq-approx} and the actual oscillation period $T$ along the cross-section (S2) as the horizontal dotted line drawn in Fig.~\ref{fig-p}, with the fixed value $a=2.2$. The simplified version of Eq.~\eqref{eq-approx} under $a = a_{12} = a_{21}$ and $r = 1$ is obtained as
\begin{eqnarray}
T_{\mathrm{homo}} = 2 \tau_y \left[ \ln \frac{1}{\delta} - \ln (1+b-a) \right]. \label{eq-approx_s}
\end{eqnarray}
The theoretical curve of Eq.~\eqref{eq-approx_s} is illustrated by the solid lines in Fig.~\ref{fig-a-period}, which better agree with the actual oscillation period $T$ (circle) than $T_{\mathrm{harm}}$ (dotted line). Note that even Eq.~\eqref{eq-approx} cannot be applied to the oscillation type (d) shown in Fig.~\ref{fig-limitcycle}(d), which has a constant oscillation period for any $r$ under fixed $a_{12}, a_{21}$.

To derive the scaling law $T_{\mathrm{homo}}$ of Eq.~\eqref{eq-approx}, we first divide an oscillation period $T$ into four pieces corresponding to the separately colored trajectories in Fig.~\ref{fig-limitcycle}(a)--(c):
\begin{eqnarray}
T = T_B + T_C + T_{D_1} + T_{D_2}, \label{eq-17}
\end{eqnarray}
where $T_B$, $T_C$, $T_{D_1}$, and $T_{D_2}$ are the time durations required to pass through $S_B$, $S_C$, $S_D$ from $\Sigma_{BD}$ to $\Sigma_{DC}$, and $S_D$ from $\Sigma_{CD}$ to $\Sigma_{DB}$, respectively.

Next, we make several main assumptions.
\begin{ass}
$T_{D_1}$ and $T_{D_2}$ can be ignored because these durations are shorter than $T_B$ and $T_C$. \label{ass1}
\end{ass}
\begin{ass}
The values of $y_1$ and $y_2$ remain almost unchanged while the system state passes the region $S_D$. \label{ass2}
\end{ass}
\begin{ass}
The system state in the region $S_B$ (or $S_C$) approaches $\bm{X}_B^*$ ($\bm{X}_C^*$) along the slowest eigenvector $\bm{v}_B$ ($\bm{v}_C$) corresponding to the eigenvalue $\lambda_B=-\frac{1}{\tau_y}$ $\left( \lambda_C=-\frac{1}{\tau_y} \right)$. \label{ass3}
\end{ass}

Applying Assumption \ref{ass1} to Eq.~\eqref{eq-17}, we obtain
\begin{eqnarray}
T \approx T_{\mathrm{homo}} = T_B + T_C. \label{eq-18}
\end{eqnarray}
According to Eq.~\eqref{eq-Ba}, the dynamics in the region $S_B$ are simply a first-order differential equation regarding $y_2$:
\begin{eqnarray}
\tau_y \frac{d y_2}{dt} = -y_2. \label{eq-dy2dt}
\end{eqnarray}
Hence, $T_B$ is simply evaluated by the variation amount of $y_2$ as
\begin{eqnarray}
T_B=\tau_y \ln \frac{ \left[\bm{X}_B^0\right]_{y_2} }{ \left[\bm{X}_B^1\right]_{y_2} }, \label{eq-19a}
\end{eqnarray}
where $\bm{X}_B^0$ and $\bm{X}_B^1$ are the initial and terminal points of the passing trajectory in $S_B$, respectively. Similarly, $y_1$ follows the dynamics
\begin{eqnarray}
\tau_y \frac{d y_1}{dt} = -y_1, \label{eq-dy1dt}
\end{eqnarray}
in the region $S_C$ considering Eq.~\eqref{eq-Ca}. Therefore, $T_C$ is given by
\begin{eqnarray}
T_C=\tau_y \ln \frac{ \left[\bm{X}_C^0\right]_{y_1} }{ \left[\bm{X}_C^1\right]_{y_1} }, \label{eq-19b}
\end{eqnarray}\label{eq-19}
where $\bm{X}_C^0$ and $\bm{X}_C^1$ are the start and end states of the oscillation path in the region $S_C$. Applying Assumption \ref{ass2} results to the following approximation:
\begin{subequations}
\begin{eqnarray}
\left[ \bm{X}_B^0 \right]_{y_2} & \approx & \left[ \bm{X}_C^1 \right]_{y_2}, \label{eq-20a} \\
\left[ \bm{X}_C^0 \right]_{y_1} & \approx & \left[ \bm{X}_B^1 \right]_{y_2}. \label{eq-20b}
\end{eqnarray}\label{eq-20}
\end{subequations}

Now, the problem is reduced to calculating $\bm{X}_B^1$ and $\bm{X}_C^1$. Considering Assumption \ref{ass3}, $\bm{X}_B^1$ is derived as an intersection between $\Sigma_{BD}$ and the slowest eigenvector $\bm{v}_B$ extended from the virtual fixed point $\bm{X}_B^*$ (this extension is roughly represented by the cyan dotted lines in Fig.~\ref{fig-limitcycle}). The eigenvector is 
\begin{eqnarray}
\bm{v}_B=
\begin{bmatrix}
0 & 0 & 1 & -\delta
\end{bmatrix}^\top. \label{eq-vB}
\end{eqnarray}
Providing that $\bm{X}_B^1 = \bm{X}_B^* + k \bm{v}_B$, then $\left[\bm{X}_B^1\right]_{x_2}=0$ when
\begin{eqnarray}
k=-\frac{(1+b)r-a_{21}}{1+b}s_1. \label{eq-k}
\end{eqnarray}
Therefore, we can confirm 
\begin{subequations}
\begin{eqnarray}
\bm{X}_B^1=
\begin{bmatrix}
\frac{s_1}{1+b} & \frac{s_1}{1+b} & 0 & \frac{(1+b)r-a_{21}}{1+b}\delta s_1
\end{bmatrix}^\top. \label{eq-21a}
\end{eqnarray}
Similarly, we get
\begin{eqnarray}
\bm{X}_C^1=
\begin{bmatrix}
0 & \frac{(1+b)-ra_{12}}{1+b}\delta s_1 & \frac{rs_1}{1+b} & \frac{rs_1}{1+b}
\end{bmatrix}^\top. \label{eq-21b}
\end{eqnarray}\label{eq-21}
\end{subequations}
Considering Eqs.~\eqref{eq-20a}, \eqref{eq-20b}, \eqref{eq-21a}, and \eqref{eq-21b}, Eqs.~\eqref{eq-19a} and \eqref{eq-19b} are written as
\begin{subequations}
\begin{eqnarray}
T_B &=& \tau_y \ln \left( \frac{1}{\delta} \frac{r}{(1+b)r-a_{21}} \right), \label{eq-22a}\\
T_C &=& \tau_y \ln \left( \frac{1}{\delta} \frac{1}{(1+b)-r a_{12}}\right). \label{eq-22b}
\end{eqnarray} \label{eq-22}
\end{subequations}
Substituting Eqs.~\eqref{eq-22a} and \eqref{eq-22b} into Eq.~\eqref{eq-18} and simplifying, the logarithmic scaling law of Eq.~\eqref{eq-approx} holds.

The order of logarithmic divergence of $T$ by Eq.~\eqref{eq-approx} indicates a ghost in the oscillation type (c), which is close to the second homoclinic-like bifurcation. When $(r,a)$ is close to the borderline $\partial \Omega_2^+$ (or $\partial \Omega_3^+$), the stable fixed point $\bm{X}_B^*$ ($\bm{X}_C^*$) is still virtual, yet it is almost on $\Sigma_{BD}$ ($\Sigma_{CD}$). Consequently, its saddle-like point effect causes a ghost, which requires a long duration to pass slowly. The orbit leading to this ghost point works as a bottleneck of time, which becomes dominant in the entire oscillation cycle.

Now we discuss the validity of the introduced approximations. Assumption \ref{ass1} serves to eliminate oscillation patterns observed when $a$ is small, such as the oscillation types (b) and (d). For oscillation types (a) and (c), this is a good approximation because the process of $T_{D_1}$ and $T_{D_2}$ is very fast. Assumption \ref{ass2} brings an even coarser approximation. For example, along the lower black trajectories in the right of Fig.~\ref{fig-limitcycle}(c), which corresponds to the duration $T_{D_2}$ of the oscillation type (c), the value of $Y = y_1-y_2$ changes sufficiently that Assumption \ref{ass2} may be inappropriate here. Another version of Eq.~\eqref{eq-approx} without Assumption \ref{ass2} is described in the Appendix.

Assumption \ref{ass3} provides a good approximation, specifically for the oscillation type (c) slightly before the homoclinic-like bifurcation, where $\bm{X}_B^*$ (or $\bm{X}_C^*$) is about to appear from the virtual state crossing the boundary $\Sigma_{BD}$ ($\Sigma_{CD}$). For example, in Fig.~\ref{fig-limitcycle}(c), the direction of the trajectory in $S_C$ (magenta solid line) in the vicinity of the virtual fixed point $\bm{X}_C^*$ (magenta-edged circle) is almost identical to the slowest eigendirection of $\bm{X}_C^*$. Here, $\bm{X}_C^*$ is positioned just before the boundary $\Sigma_{CD}$ is crossed. This approximation is even worse for the oscillation type (b) shown in Fig.~\ref{fig-limitcycle}(b). According to the formulation in \eqref{eq-Bb} [or \eqref{eq-Cb}], the $x_2$-coordinate of $\bm{X}_B^*$ ($x_1$-coordinate of $\bm{X}_C^*$) is expected to increase as $a_{21}$ ($a_{12}$) decreases; the oscillation pattern approaches (b); thus, $\bm{X}_B^*$ ($\bm{X}_C^*$) moves further away from $\Sigma_{BD}$ ($\Sigma_{CD}$). According to the numerically simulated trajectories in Fig.~\ref{fig-limitcycle}, the longer distance between $\bm{X}_B^*$ ($\bm{X}_C^*$) and $\Sigma_{BD}$ ($\Sigma_{CD}$) is responsible for the inappropriate approximation by Assumption \ref{ass3}.
 
\subsection{Noise-induced Oscillation \label{subsec5D}}

A novel result predicted regarding the second homoclinic-like bifurcation is that noise-induced oscillations emerge even through the parameter set $(r, a_{12}, a_{21})$ is on or slightly outside $\partial \Omega_2$ (or $\partial \Omega_3$). In this condition, the original noiseless system \eqref{eq-1} only converges to the stable fixed point $\bm{X}_B^*$ ($\bm{X}_C^*$) existing in the vicinity of $\Sigma_{BD}$ ($\Sigma_{CD}$). This process is represented by the activity pattern of Fig.~\ref{fig-t-xyz}(f). To introduce external noise to the system, for example, we can replace Eq.~\eqref{eq-1a} by
\begin{eqnarray}
\tau_x \frac{dx_i}{dt} = -x_i-by_i - \sum_{j \neq i}^{n} a_{ij} z_j + s_i + \sigma \eta, \label{eq-noise}
\end{eqnarray}
where $\sigma$ determines the noise intensity and $\eta$ is the standard Gaussian noise.

\begin{figure*}
\includegraphics[width=\linewidth]{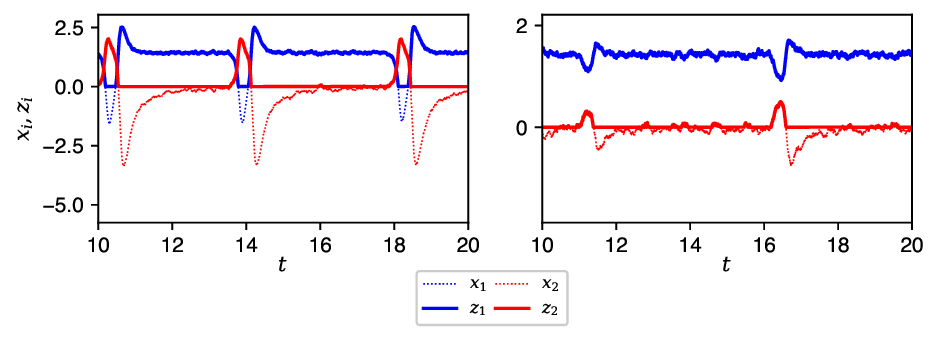}
\caption{Neuronal activity patterns observed in the noisy system \eqref{eq-3} with Eq.~\eqref{eq-3a} replaced by Eq.~\eqref{eq-noise}, where $\sigma = 0.2$ and $a=a_{12}=a_{21}$; (Left) $a=2$, $r=0.56$, which is the same condition as Fig.~\ref{fig-t-xyz}(f). Similar to oscillation type (c). (Right) $a=1.6$, $r=0.455$. Similar to oscillation type (d).}
\label{fig-t-xz_noise}
\end{figure*}

If the noise intensity $\sigma$ is sufficient, then the system state which almost converges to $\bm{X}_B^*$ ($\bm{X}_C^*$) may by chance jump across $\Sigma_{BD}$ ($\Sigma_{CD}$) into $S_D^* \subset S_D$. Here, $S_D^*$ is a subregion in which there is a locally attracting homoclinic-like orbit inherited from the periodic orbits observed when $(r, a_{12}, a_{21})$ is inside $\partial \Omega_2$ ($\partial \Omega_3$). Nearly riding on this homoclinic-like orbit, the system state in turn starts an excursion through $S_D$, $S_C$, again $S_D$, and $S_B$ [corresponding to the oscillation pattern (c)], or only through $S_D$ and $S_B$ [oscillation pattern (d)], returning to $\bm{X}_B^*$ ($\bm{X}_C^*$). This is the overall dynamical picture of the noise-induced oscillations. Figure~\ref{fig-t-xz_noise} shows examples of noise-induced oscillations observed in the vicinity of the homoclinic-like and supercritical Hopf-like bifurcations, the waveforms of which are inherited from the (c) and (d) oscillation patterns.

More practically, a noise-induced oscillation cycle can be defined as an itinerary that starts upon crossing $\Sigma_{BD}$ ($\Sigma_{CD}$) into $S_D^*$, undergoes a certain excursion returning to $S_B$ ($S_C$), and terminates on crossing $\Sigma_{BD}$ ($\Sigma_{CD}$) again. We can also define the oscillation period of a noise-induced oscillation as $T_\sigma$, which is the duration required for such a cycle. Here, $T_\sigma$ is stochastically variable. When the noise is sufficiently small, the excursion process occupies only a small proportion of $T_\sigma$, and most of the remaining duration is allocated to the process of jumping from $\bm{X}_B^*$ ($\bm{X}_C^*$) across $\Sigma_{BD}$ ($\Sigma_{CD}$) into $S_D^*$ which is driven by the external noise perturbations. This jumping duration is estimated as the minimum time spent to reach $S_D^*$ and is expected to be shorter as the stochastic perturbation gets larger on average. In addition, $\bm{X}_B^*$ ($\bm{X}_C^*$) separates further from $\Sigma_{BD}$ ($\Sigma_{CD}$) as the parameter set $(r, a_{12}, a_{21})$ are distanced from the borderline $\partial \Omega_2$ or $\partial \Omega_3$. Under the same noise intensity $\sigma$, the duration required for the jumping process could increase as the jumping distance extends.

\begin{figure}
\includegraphics[width=\linewidth]{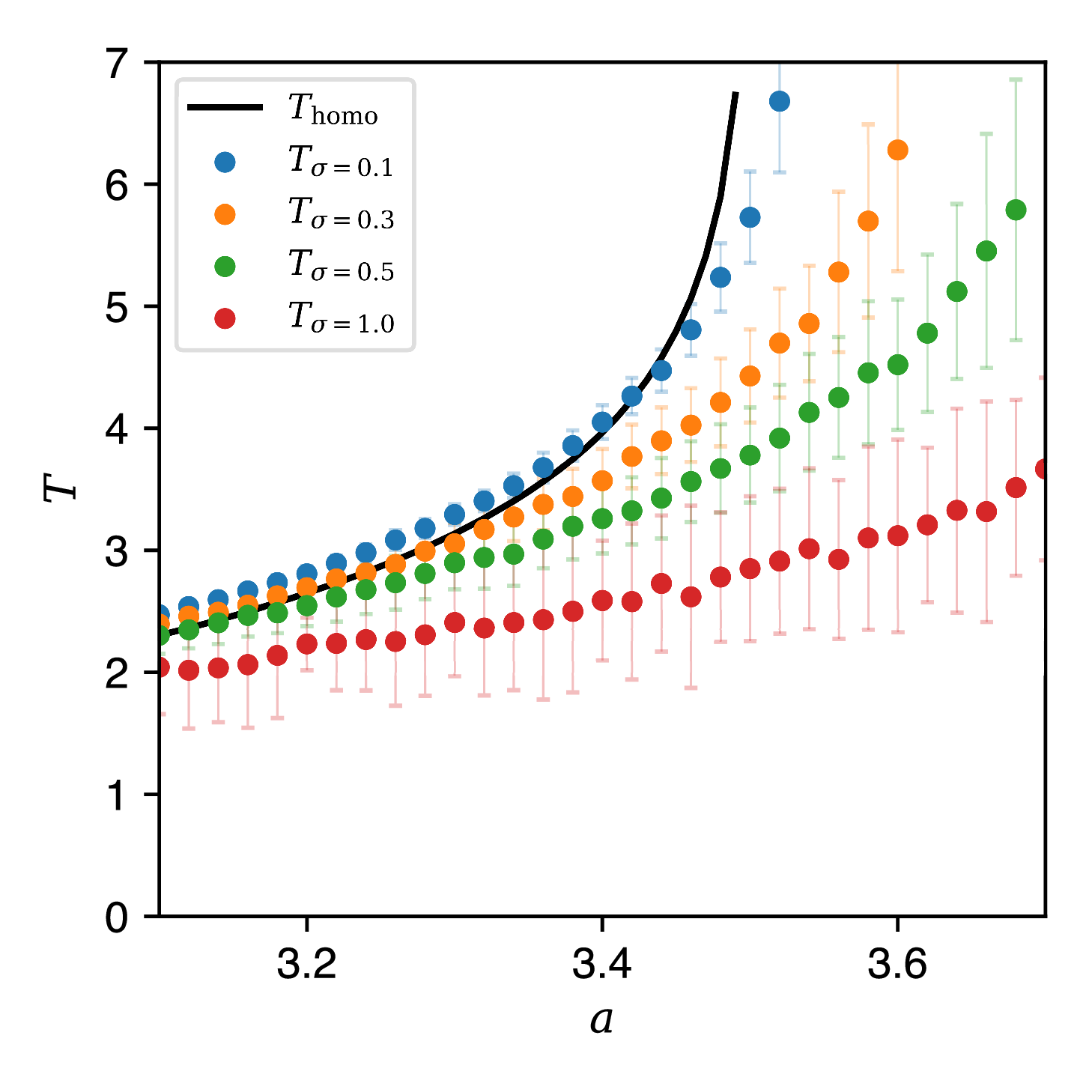}
\caption{Plot of average periods $T_\sigma$ of noise-induced oscillations for various $\sigma$ and $a=a_{12}=a_{21}$ under the condition $r=1$, with the plot of $T_{\mathrm{homo}}$ (black) in Fig.~\ref{fig-a-period}. Each circle plot is calculated by averaging the periods of 200 numerically simulated noise-induced oscillation cycles, with error bars of their standard deviations.}
\label{fig-a-period_noise}
\end{figure}

Thus, $T_\sigma$ could increase as the noise intensity $\sigma$ declines, or the parameter set $(r, a_{12}, a_{21})$ distances from the borderline $\partial \Omega_2$ or $\partial \Omega_3$. From another perspective, the divergence of $T_\sigma$ is relieved and the oscillation-existing area $\Omega$ in the parameter space is widened, for a larger $\sigma$. These predictions can be numerically verified by evaluating the arithmetic mean of $T_\sigma$ for multiple oscillation cycles (Fig.~\ref{fig-a-period_noise}).

\section{DISCUSSION \label{sec6}}

This study's first step was the fixed point analysis of the two-neuron Matsuoka oscillator model (Section~\ref{sec4}), which enabled a quantitative evaluation of transient neuronal activities as convergences to stable fixed points. We found that oscillatory solutions are caused by convergence to the virtual stable fixed points $\bm{X}_B^*$ and $\bm{X}_C^*$ and divergence from the existing unstable fixed point $\bm{X}_D^*$ (Subsection~\ref{subsec5A}). This could be why the conditions for absence of the stable fixed points are equivalent to the existence condition of oscillations. Furthermore, this oscillation picture provided explanations of several bifurcation scenarios between oscillatory and stationary states, or between different oscillation types: subcritical Hopf-like, homoclinic-like, supercritical Hopf-like, and grazing bifurcations (Subsection~\ref{subsec5B}). The additional knowledge presented in this paper such as the logarithmic oscillation-period scaling law (Subsection~\ref{subsec5C}) and noise-induced oscillation (Subsection~\ref{subsec5D}) were also obtained following the fixed point analysis in Section~\ref{sec4}.

\subsection{Mathematical Aspects \label{subsec6A}}

The logarithmic scaling law \eqref{eq-approx}, as a new approximation $T_{\mathrm{homo}}$ for the oscillation period $T$, was derived in terms of the convergent linear dynamics of the variables $y_1$ and $y_2$ in the regions $S_B$ and $S_C$. The order of the divergence of $T_{\mathrm{homo}}$ in this scaling law is similarly observed in the neighborhood of homoclinic bifurcations.\cite{gaspard1990measurement, santos2004effects} However, the ``homoclinic-like'' bifurcation discussed in Subsection~\ref{subsec5B} concerning this scaling law is slightly different from other homoclinic bifurcations. Standard homoclinic bifurcation has a saddle point approaching a limit cycle orbit. When these collide and fuse, the saddle point rides on the limit cycle orbit with the periodic oscillatory nature disappearing. Conversely, this model only has a virtual stable node $\bm{X}_B^*$ (or $\bm{X}_C^*$) instead of a saddle point. At the homoclinic-like bifurcation moment, $\bm{X}_B^*$ ($\bm{X}_C^*$) reaches the boundary $\Sigma_{BD}$ ($\Sigma_{CD}$) and transforms from a virtual into a regular state. Because oscillations are caused by convergence to virtual $\bm{X}_B^*$ ($\bm{X}_C^*$), at this homoclinic-like bifurcation, it naturally emerges on the original periodic orbit. Moreover, in this situation, the dynamics flow into the $\bm{X}_B^*$ ($\bm{X}_C^*$) from the region $S_B$ ($S_C$) and escape to $S_D$, which made $\bm{X}_B^*$ ($\bm{X}_C^*$) assume a saddle-like property.

Beyond this bifurcation, $\bm{X}_B^*$ ($\bm{X}_C^*$) distances from $\Sigma_{BD}$ ($\Sigma_{CD}$). Even in this situation, an oscillatory excursion along the homoclinic-like orbit could occur if a sufficient perturbation is adopted to cross over that gap. This is the noise-induced oscillation proposed in Subection~\ref{subsec5D}, which is essentially common to phenomena described as ``noise-invoked resonance,'' for instance, in previous studies concerning homoclinic bifurcations.\cite{santos2004effects, nurujjaman2008noise} It is plausible that as biological systems, neural circuits are always influenced by noise. Therefore, the results concerning the noise-induced oscillation in the present study would provide a foundation for validating CPG models through biological experiments.

Unlike the logarithmic approximation $T_{\mathrm{homo}}$, the previously proposed approximation $T_{\mathrm{harm}}$ given by Eq.~\eqref{eq-matsuoka2011} was derived in terms of an approximation of a periodic cycle into a behavior of the harmonic oscillator in the region $S_D$. In particular, at $\partial \Omega_1$, this approximation is strictly accurate because the oscillation pattern becomes neutrally stable around a center, $\bm{X}_D^*$. This change in the stability of $\bm{X}_D^*$ is responsible for the subcritical Hopf-like bifurcation in the vicinity of the borderline $\partial \Omega_1$. There remains the problem that the previous approximation $T_{\mathrm{harm}}$ is valid only in the symmetric network $a=a_{12}=a_{21}$ and not applicable to asymmetric cases $a_{12} \neq a_{21}$ like the new approximation $T_{\mathrm{homo}}$. Hence, it is a future prospect to modify $T_{\mathrm{harm}}$ given by Eq.~\eqref{eq-matsuoka2011} into more general formulation which also covers asymmetric cases. This approach may be even essential to answer another question of why the oscillation type (d) has an invariant oscillation period independent of the value of $r$.

The oscillation types (c) and (d) were found by eliminating the assumption of symmetry input stimuli into the two neurons ($r=1$), which has been adopted in many previous studies. For example, Ref.~\onlinecite{gonccalves2005regions} investigated the stability of a limit-cycle solution in the symmetric $r=1$ case of the two-neuron Matsuoka oscillator model. Because the stability of the oscillation types (c) and (d) was only numerically confirmed in this paper and not theoretically guaranteed, a future undertaking is to resolve this problem by extending the method of impact maps described in Ref.~\onlinecite{gonccalves2005regions}.

\subsection{Biological Interpretations and Limitations \label{subsec6B}}

The external inputs into neurons $s_1$ and $s_2$ were set as constants throughout this study. Suppose that values of the synaptic weights $a_{12}$ and $a_{21}$ are fixed to satisfy Ineq.~\eqref{eq-add1}, the input ratio $r=\frac{s_2}{s_1}$ is the only bifurcation parameter between stable stationary states and stable oscillations. Gradual and continuous incrementation of $r$ is equal to alterations along the horizontal axis like (S2) in Fig.~\ref{fig-p}, which contributes to phase transition among
\begin{itemize}
\setlength{\leftskip}{10pt }
\item convergent dynamics to $\bm{X}_B^*$ when \eqref{eq-Bd},
\item oscillatory dynamics under the condition \eqref{eq-Dd}, and
\item convergent dynamics to $\bm{X}_C^*$ if \eqref{eq-Cd} is true.
\end{itemize}
In this sense, $\bm{X}_B^*$ or $\bm{X}_C^*$ are supposed to be the start or end points of transient neuronal activities. This could be a hypothesis for the continuous transition between oscillatory and convergent neuronal activities, and even between rhythmic and discrete movements.

The present analysis suggests that the Matsuoka oscillator model has the restriction that $\bm{X}_D^*$ cannot be a stable fixed point when the values of the synaptic weights $a_{12}$, $a_{21}$ are fixed to satisfy the existence condition of oscillatory solutions \eqref{eq-add1}. Two coupled neurons of the Matsuoka oscillator model have been used to model the flexion and extension of a single joint by matching each neuron to either a flexor or extensor muscle.\cite{de2003interaction} In this situation, however, it is impossible to achieve the stationary states where the two neurons are simultaneously activated leading to stationary co-contraction of the muscles. To reproduce this sustainable co-activation of neurons with the original Matsuoka oscillator model, synaptic weights $a_{12}$ and $a_{21}$ should change as fast as the ratio of inputs into the neurons $r=\frac{s_2}{s_1}$. One possible scheme is to assume specific interneurons between the two main neurons. Parameters $a_{12}$ and $a_{21}$ are regarded as the overall transmission efficiency between the two main neurons, mediated by the interneurons, the values of which are regulated by other external input signals imposed on the interneurons. This structure should involve a time delay in the process of multiple synaptic transmissions, although this is not assumed in the original Matsuoka oscillator model. Improving the original model into a more plausible one that is consistent with physiological findings on neurons and neuronal networks will be an important future development.

Finally, the analysis of the Matsuoka oscillator model in this study was consistently limited to the two-neuron case. When the number of neurons is $n \geq 3$, the network topologies become more complex and analysis is harder because it requires the solving of sixth- or higher-order equations. However, the fixed point analysis in this study may be useful in understanding the properties of the multi-neuron dynamical system. Because there are $n$ variables of the membrane potential $x_i \ (i=1,2,\cdots,n)$, the solution space is divided into $2^n$ regions regarding the positivity or negativity of $x_i$. Thus, we  could investigate the dynamics specific to each region, the fixed points in the relevant dynamics, and the existence and stability of the fixed points. Extending the present results to the general $n$-neuron version of the Mastuoka oscillator model may be necessary to understand CPG circuits in the spinal cord as a huge multidimensional system.

\begin{acknowledgments}
We thank the lab members in Nonlinear Physics Group (Kori-Kobayashi-Izumida Group) and peers in The Japanese Society for Motor Control and The Society for the Neural Control of Movement, for research discussion and constructive criticism. This study was supported by JSPS KAKENHI (No. JP24KJ0876) and Scholarship for Graduate Students by Nakatani Foundation for Advancement of Measuring Technologies in Biomedical Engineering to K.M., and JSPS KAKENHI (No. JP21J10799, JP23H02796) to H.K.
\end{acknowledgments}

\section*{AUTHOR DECLARATIONS}

\subsection*{Conflict of Interest}
The authors have no conflicts to disclose.

\subsection*{Author Contributions}
K.M and H.K. conceived the project. K.M. performed theoretical and numerical analysis. K.M. wrote the manuscript under the supervision by H.K.

\section*{Data Availability Statement}
The data that support the findings of this study are available within the article [and its supplementary material].

\appendix*
\section{Revised Version of the Logarithmic Oscillation-Period Scaling Law \label{secapp}}
\begin{figure*}
 \centering
 \includegraphics[width=1\linewidth]{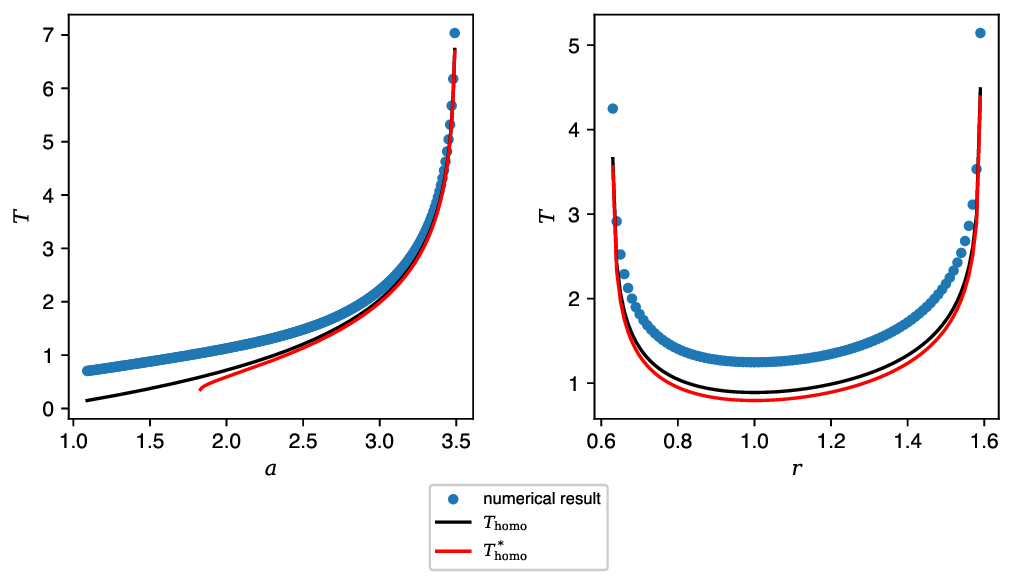}
 \caption{The curves of the revised version $T_{\mathrm{homo}}^*$ are added to the left-hand side of Fig.~\ref{fig-a-period} (Left), and the left-hand side of Fig.~\ref{fig-a-r-p} (Right). Note that $T_{\mathrm{homo}}^*$ is only defined in the $a \geq a_+^*$ region.}
 \label{fig-period-appx}
\end{figure*}

Instead of Assumption \ref{ass2}, a new one could be introduced;
\begin{ass}
The values of $y_1$ and $y_2$ change along the fastest eigenvector $\bm{v}_D$ of the unstable fixed point $\bm{X}_D^*$ in region $S_D$. \label{ass4}
\end{ass}
This can be available only when $\bm{X}_D^*$ is an unstable node, not an unstable spiral, which occurs under the $a \geq a_{+}^*$ condition. Additionally, Assumption \ref{ass4} is consistent with Assumption \ref{ass1}; thus, the oscillation types (a) and (c) also serve for good approximation by Assumption \ref{ass4} because under this condition, the diverging effect of the slower eigenvector reduces throughout $T_{D_1}$ and $T_{D_2}$. The fastest eigenvector $\bm{v}_D$ is
\begin{eqnarray}
\bm{v}_D=
\begin{bmatrix}
1 & v & -\alpha & -\alpha v
\end{bmatrix}^\top, \label{eq-vD}
\end{eqnarray}
where
\begin{eqnarray}
\alpha &=& \sqrt{ \frac{a_{21}}{a_{12}} }, \label{eq-alpha}\\
v &=& \frac{\tau_x-(1-\sqrt{a_{12}a_{21}})\tau_y-\sqrt{Q_{D-}}}{2b\tau_y}. \label{eq-v}
\end{eqnarray}
Note that $Q_{D-}$ is the discriminant of Eq.~\eqref{eq-D-}. Instead of Eqs.~\eqref{eq-20a} and \eqref{eq-20b}, Assumption \ref{ass4} leads to 
\begin{subequations}
\begin{eqnarray}
\bm{X}_B^0 = \bm{X}_C^1 + k_B \bm{v}_D, \label{eq-X_B^0}\\
\bm{X}_C^0 = \bm{X}_B^1 + k_C \bm{v}_D. \label{eq-X_C^0}
\end{eqnarray} \label{eq-app1}
\end{subequations}
Solving $\left[\bm{X}_B^0\right]_{x_2}=0$ and $\left[\bm{X}_C^0\right]_{x_1}=0$, we obtain
\begin{subequations}
\begin{eqnarray}
k_B = \frac{rs_1}{\alpha(1+b)}, \label{eq-k_B}\\
k_C = -\frac{s_1}{1+b}, \label{eq-k_C}
\end{eqnarray} \label{eq-app1}
\end{subequations}
which is followed by
\begin{subequations}
\begin{eqnarray}
\left[\bm{X}_B^0\right]_{y_2} &=& (1-v)\frac{r s_1}{1+b}, \label{eq-app1a}\\
\left[\bm{X}_C^0\right]_{y_1} &=& (1-v)\frac{s_1}{1+b}. \label{eq-app1b}
\end{eqnarray} \label{eq-app1}
\end{subequations}
Substituting Eqs.~\eqref{eq-21a}, \eqref{eq-21b}, \eqref{eq-app1a}, and \eqref{eq-app1b} into Eqs.~\eqref{eq-19a} and \eqref{eq-19b}, we get
\begin{subequations}
\begin{eqnarray}
T_B &=& \tau_y \ln \left( \frac{1}{\delta} \frac{(1-v)r}{(1+b)r-a_{21}} \right), \label{eq-app2a}\\
T_C &=& \tau_y \ln \left( \frac{1}{\delta} \frac{(1-v)1}{(1+b)-r a_{12}}\right). \label{eq-app2b}
\end{eqnarray} \label{eq-app2}
\end{subequations}
Finally, in applying Eqs.~\eqref{eq-app2a} and \eqref{eq-app2b} to Eq.~\eqref{eq-18}, the following result holds:
\begin{eqnarray}
T_{\mathrm{homo}}^* = \tau_y \left[ \ln \frac{(1-v)^2}{(1+b)a_{12} \delta^2} - \ln \frac{(r_{\mathrm{sup}}-r)(r-r_{\mathrm{inf}})}{r} \right]. \label{eq-approx_rev}
\end{eqnarray}
This is the revised version of the oscillation-period scaling law. Compared with the original version, $T_{\mathrm{homo}}$, given by \eqref{eq-approx}, this form $T_{\mathrm{homo}}^*$ differs only in the constant term for the coefficient $(1-v)^2$ in the logarithm. 
Figure~\ref{fig-period-appx} shows the curve of $T_{\mathrm{homo}}^*$ in the same form as Figs.~\ref{fig-a-period} and \ref{fig-a-r-p}. Here, $T_{\mathrm{homo}}^* \leq T_{\mathrm{homo}}$ holds so that $T_{\mathrm{homo}}^*$ seems less accurate than $T_{\mathrm{homo}}$ for the numerically observed oscillation period $T$. This is because the duration $T_D$ is ignored according to Assumption \ref{ass1}. If $T_D$ is accurately reckoned, then $T_{\mathrm{homo}}^*$ will be closer to $T$ than $T_{\mathrm{homo}}$.

\nocite{*}
\bibliography{muramatsu20240514}

\providecommand{\noopsort}[1]{}\providecommand{\singleletter}[1]{#1}
\begin{thebibliography}{32}%
\makeatletter
\providecommand \@ifxundefined [1]{%
 \@ifx{#1\undefined}
}%
\providecommand \@ifnum [1]{%
 \ifnum #1\expandafter \@firstoftwo
 \else \expandafter \@secondoftwo
 \fi
}%
\providecommand \@ifx [1]{%
 \ifx #1\expandafter \@firstoftwo
 \else \expandafter \@secondoftwo
 \fi
}%
\providecommand \natexlab [1]{#1}%
\providecommand \enquote  [1]{``#1''}%
\providecommand \bibnamefont  [1]{#1}%
\providecommand \bibfnamefont [1]{#1}%
\providecommand \citenamefont [1]{#1}%
\providecommand \href@noop [0]{\@secondoftwo}%
\providecommand \href [0]{\begingroup \@sanitize@url \@href}%
\providecommand \@href[1]{\@@startlink{#1}\@@href}%
\providecommand \@@href[1]{\endgroup#1\@@endlink}%
\providecommand \@sanitize@url [0]{\catcode `\\12\catcode `\$12\catcode
  `\&12\catcode `\#12\catcode `\^12\catcode `\_12\catcode `\%12\relax}%
\providecommand \@@startlink[1]{}%
\providecommand \@@endlink[0]{}%
\providecommand \url  [0]{\begingroup\@sanitize@url \@url }%
\providecommand \@url [1]{\endgroup\@href {#1}{\urlprefix }}%
\providecommand \urlprefix  [0]{URL }%
\providecommand \Eprint [0]{\href }%
\providecommand \doibase [0]{http://dx.doi.org/}%
\providecommand \selectlanguage [0]{\@gobble}%
\providecommand \bibinfo  [0]{\@secondoftwo}%
\providecommand \bibfield  [0]{\@secondoftwo}%
\providecommand \translation [1]{[#1]}%
\providecommand \BibitemOpen [0]{}%
\providecommand \bibitemStop [0]{}%
\providecommand \bibitemNoStop [0]{.\EOS\space}%
\providecommand \EOS [0]{\spacefactor3000\relax}%
\providecommand \BibitemShut  [1]{\csname bibitem#1\endcsname}%
\let\auto@bib@innerbib\@empty
\bibitem [{\citenamefont {Ijspeert}\ \emph {et~al.}(2007)\citenamefont
  {Ijspeert}, \citenamefont {Crespi}, \citenamefont {Ryczko},\ and\
  \citenamefont {Cabelguen}}]{ijspeert2007swimming}%
  \BibitemOpen
  \bibfield  {author} {\bibinfo {author} {\bibfnamefont {A.~J.}\ \bibnamefont
  {Ijspeert}}, \bibinfo {author} {\bibfnamefont {A.}~\bibnamefont {Crespi}},
  \bibinfo {author} {\bibfnamefont {D.}~\bibnamefont {Ryczko}}, \ and\ \bibinfo
  {author} {\bibfnamefont {J.-M.}\ \bibnamefont {Cabelguen}},\ }\bibfield
  {title} {\enquote {\bibinfo {title} {From swimming to walking with a
  salamander robot driven by a spinal cord model},}\ }\href@noop {} {\bibfield
  {journal} {\bibinfo  {journal} {science}\ }\textbf {\bibinfo {volume}
  {315}},\ \bibinfo {pages} {1416--1420} (\bibinfo {year} {2007})}\BibitemShut
  {NoStop}%
\bibitem [{\citenamefont {Ijspeert}(2008)}]{ijspeert2008central}%
  \BibitemOpen
  \bibfield  {author} {\bibinfo {author} {\bibfnamefont {A.~J.}\ \bibnamefont
  {Ijspeert}},\ }\bibfield  {title} {\enquote {\bibinfo {title} {Central
  pattern generators for locomotion control in animals and robots: a review},}\
  }\href@noop {} {\bibfield  {journal} {\bibinfo  {journal} {Neural networks}\
  }\textbf {\bibinfo {volume} {21}},\ \bibinfo {pages} {642--653} (\bibinfo
  {year} {2008})}\BibitemShut {NoStop}%
\bibitem [{\citenamefont {Grillner}(2006)}]{grillner2006biological}%
  \BibitemOpen
  \bibfield  {author} {\bibinfo {author} {\bibfnamefont {S.}~\bibnamefont
  {Grillner}},\ }\bibfield  {title} {\enquote {\bibinfo {title} {Biological
  pattern generation: the cellular and computational logic of networks in
  motion},}\ }\href@noop {} {\bibfield  {journal} {\bibinfo  {journal}
  {Neuron}\ }\textbf {\bibinfo {volume} {52}},\ \bibinfo {pages} {751--766}
  (\bibinfo {year} {2006})}\BibitemShut {NoStop}%
\bibitem [{\citenamefont {Takei}\ and\ \citenamefont
  {Seki}(2013)}]{takei2013spinal}%
  \BibitemOpen
  \bibfield  {author} {\bibinfo {author} {\bibfnamefont {T.}~\bibnamefont
  {Takei}}\ and\ \bibinfo {author} {\bibfnamefont {K.}~\bibnamefont {Seki}},\
  }\bibfield  {title} {\enquote {\bibinfo {title} {Spinal premotor interneurons
  mediate dynamic and static motor commands for precision grip in monkeys},}\
  }\href@noop {} {\bibfield  {journal} {\bibinfo  {journal} {Journal of
  Neuroscience}\ }\textbf {\bibinfo {volume} {33}},\ \bibinfo {pages}
  {8850--8860} (\bibinfo {year} {2013})}\BibitemShut {NoStop}%
\bibitem [{\citenamefont {Azim}\ \emph {et~al.}(2014)\citenamefont {Azim},
  \citenamefont {Jiang}, \citenamefont {Alstermark},\ and\ \citenamefont
  {Jessell}}]{azim2014skilled}%
  \BibitemOpen
  \bibfield  {author} {\bibinfo {author} {\bibfnamefont {E.}~\bibnamefont
  {Azim}}, \bibinfo {author} {\bibfnamefont {J.}~\bibnamefont {Jiang}},
  \bibinfo {author} {\bibfnamefont {B.}~\bibnamefont {Alstermark}}, \ and\
  \bibinfo {author} {\bibfnamefont {T.~M.}\ \bibnamefont {Jessell}},\
  }\bibfield  {title} {\enquote {\bibinfo {title} {Skilled reaching relies on a
  v2a propriospinal internal copy circuit},}\ }\href@noop {} {\bibfield
  {journal} {\bibinfo  {journal} {Nature}\ }\textbf {\bibinfo {volume} {508}},\
  \bibinfo {pages} {357--363} (\bibinfo {year} {2014})}\BibitemShut {NoStop}%
\bibitem [{\citenamefont {Matsuoka}(1985)}]{matsuoka1985sustained}%
  \BibitemOpen
  \bibfield  {author} {\bibinfo {author} {\bibfnamefont {K.}~\bibnamefont
  {Matsuoka}},\ }\bibfield  {title} {\enquote {\bibinfo {title} {Sustained
  oscillations generated by mutually inhibiting neurons with adaptation},}\
  }\href@noop {} {\bibfield  {journal} {\bibinfo  {journal} {Biological
  cybernetics}\ }\textbf {\bibinfo {volume} {52}},\ \bibinfo {pages} {367--376}
  (\bibinfo {year} {1985})}\BibitemShut {NoStop}%
\bibitem [{\citenamefont {Matsuoka}(1987)}]{matsuoka1987mechanisms}%
  \BibitemOpen
  \bibfield  {author} {\bibinfo {author} {\bibfnamefont {K.}~\bibnamefont
  {Matsuoka}},\ }\bibfield  {title} {\enquote {\bibinfo {title} {Mechanisms of
  frequency and pattern control in the neural rhythm generators},}\ }\href@noop
  {} {\bibfield  {journal} {\bibinfo  {journal} {Biological cybernetics}\
  }\textbf {\bibinfo {volume} {56}},\ \bibinfo {pages} {345--353} (\bibinfo
  {year} {1987})}\BibitemShut {NoStop}%
\bibitem [{\citenamefont {Matsuoka}(2011)}]{matsuoka2011analysis}%
  \BibitemOpen
  \bibfield  {author} {\bibinfo {author} {\bibfnamefont {K.}~\bibnamefont
  {Matsuoka}},\ }\bibfield  {title} {\enquote {\bibinfo {title} {Analysis of a
  neural oscillator},}\ }\href@noop {} {\bibfield  {journal} {\bibinfo
  {journal} {Biological cybernetics}\ }\textbf {\bibinfo {volume} {104}},\
  \bibinfo {pages} {297--304} (\bibinfo {year} {2011})}\BibitemShut {NoStop}%
\bibitem [{\citenamefont {Angelidis}\ \emph {et~al.}(2021)\citenamefont
  {Angelidis}, \citenamefont {Buchholz}, \citenamefont {Arreguit},
  \citenamefont {Roug{\'e}}, \citenamefont {Stewart}, \citenamefont {von
  Arnim}, \citenamefont {Knoll},\ and\ \citenamefont
  {Ijspeert}}]{angelidis2021spiking}%
  \BibitemOpen
  \bibfield  {author} {\bibinfo {author} {\bibfnamefont {E.}~\bibnamefont
  {Angelidis}}, \bibinfo {author} {\bibfnamefont {E.}~\bibnamefont {Buchholz}},
  \bibinfo {author} {\bibfnamefont {J.}~\bibnamefont {Arreguit}}, \bibinfo
  {author} {\bibfnamefont {A.}~\bibnamefont {Roug{\'e}}}, \bibinfo {author}
  {\bibfnamefont {T.}~\bibnamefont {Stewart}}, \bibinfo {author} {\bibfnamefont
  {A.}~\bibnamefont {von Arnim}}, \bibinfo {author} {\bibfnamefont
  {A.}~\bibnamefont {Knoll}}, \ and\ \bibinfo {author} {\bibfnamefont
  {A.}~\bibnamefont {Ijspeert}},\ }\bibfield  {title} {\enquote {\bibinfo
  {title} {A spiking central pattern generator for the control of a simulated
  lamprey robot running on spinnaker and loihi neuromorphic boards},}\
  }\href@noop {} {\bibfield  {journal} {\bibinfo  {journal} {Neuromorphic
  Computing and Engineering}\ }\textbf {\bibinfo {volume} {1}},\ \bibinfo
  {pages} {014005} (\bibinfo {year} {2021})}\BibitemShut {NoStop}%
\bibitem [{\citenamefont {Taga}, \citenamefont {Yamaguchi},\ and\ \citenamefont
  {Shimizu}(1991)}]{taga1991self}%
  \BibitemOpen
  \bibfield  {author} {\bibinfo {author} {\bibfnamefont {G.}~\bibnamefont
  {Taga}}, \bibinfo {author} {\bibfnamefont {Y.}~\bibnamefont {Yamaguchi}}, \
  and\ \bibinfo {author} {\bibfnamefont {H.}~\bibnamefont {Shimizu}},\
  }\bibfield  {title} {\enquote {\bibinfo {title} {Self-organized control of
  bipedal locomotion by neural oscillators in unpredictable environment},}\
  }\href@noop {} {\bibfield  {journal} {\bibinfo  {journal} {Biological
  cybernetics}\ }\textbf {\bibinfo {volume} {65}},\ \bibinfo {pages} {147--159}
  (\bibinfo {year} {1991})}\BibitemShut {NoStop}%
\bibitem [{\citenamefont {Taga}(1995)}]{taga1995model}%
  \BibitemOpen
  \bibfield  {author} {\bibinfo {author} {\bibfnamefont {G.}~\bibnamefont
  {Taga}},\ }\bibfield  {title} {\enquote {\bibinfo {title} {A model of the
  neuro-musculo-skeletal system for human locomotion: I. emergence of basic
  gait},}\ }\href@noop {} {\bibfield  {journal} {\bibinfo  {journal}
  {Biological cybernetics}\ }\textbf {\bibinfo {volume} {73}},\ \bibinfo
  {pages} {97--111} (\bibinfo {year} {1995})}\BibitemShut {NoStop}%
\bibitem [{\citenamefont {Ogihara}\ and\ \citenamefont
  {Yamazaki}(2001)}]{ogihara2001generation}%
  \BibitemOpen
  \bibfield  {author} {\bibinfo {author} {\bibfnamefont {N.}~\bibnamefont
  {Ogihara}}\ and\ \bibinfo {author} {\bibfnamefont {N.}~\bibnamefont
  {Yamazaki}},\ }\bibfield  {title} {\enquote {\bibinfo {title} {Generation of
  human bipedal locomotion by a bio-mimetic neuro-musculo-skeletal model},}\
  }\href@noop {} {\bibfield  {journal} {\bibinfo  {journal} {Biological
  cybernetics}\ }\textbf {\bibinfo {volume} {84}},\ \bibinfo {pages} {1--11}
  (\bibinfo {year} {2001})}\BibitemShut {NoStop}%
\bibitem [{\citenamefont {Wilson}\ and\ \citenamefont
  {Cowan}(1972)}]{wilson1972excitatory}%
  \BibitemOpen
  \bibfield  {author} {\bibinfo {author} {\bibfnamefont {H.~R.}\ \bibnamefont
  {Wilson}}\ and\ \bibinfo {author} {\bibfnamefont {J.~D.}\ \bibnamefont
  {Cowan}},\ }\bibfield  {title} {\enquote {\bibinfo {title} {Excitatory and
  inhibitory interactions in localized populations of model neurons},}\
  }\href@noop {} {\bibfield  {journal} {\bibinfo  {journal} {Biophysical
  journal}\ }\textbf {\bibinfo {volume} {12}},\ \bibinfo {pages} {1--24}
  (\bibinfo {year} {1972})}\BibitemShut {NoStop}%
\bibitem [{\citenamefont {Wilson}\ and\ \citenamefont
  {Cowan}(1973)}]{wilson1973mathematical}%
  \BibitemOpen
  \bibfield  {author} {\bibinfo {author} {\bibfnamefont {H.~R.}\ \bibnamefont
  {Wilson}}\ and\ \bibinfo {author} {\bibfnamefont {J.~D.}\ \bibnamefont
  {Cowan}},\ }\bibfield  {title} {\enquote {\bibinfo {title} {A mathematical
  theory of the functional dynamics of cortical and thalamic nervous tissue},}\
  }\href@noop {} {\bibfield  {journal} {\bibinfo  {journal} {Kybernetik}\
  }\textbf {\bibinfo {volume} {13}},\ \bibinfo {pages} {55--80} (\bibinfo
  {year} {1973})}\BibitemShut {NoStop}%
\bibitem [{\citenamefont {Degallier}\ and\ \citenamefont
  {Ijspeert}(2010)}]{degallier2010modeling}%
  \BibitemOpen
  \bibfield  {author} {\bibinfo {author} {\bibfnamefont {S.}~\bibnamefont
  {Degallier}}\ and\ \bibinfo {author} {\bibfnamefont {A.}~\bibnamefont
  {Ijspeert}},\ }\bibfield  {title} {\enquote {\bibinfo {title} {Modeling
  discrete and rhythmic movements through motor primitives: a review},}\
  }\href@noop {} {\bibfield  {journal} {\bibinfo  {journal} {Biological
  cybernetics}\ }\textbf {\bibinfo {volume} {103}},\ \bibinfo {pages}
  {319--338} (\bibinfo {year} {2010})}\BibitemShut {NoStop}%
\bibitem [{\citenamefont {de~Rugy}\ and\ \citenamefont
  {Sternad}(2003)}]{de2003interaction}%
  \BibitemOpen
  \bibfield  {author} {\bibinfo {author} {\bibfnamefont {A.}~\bibnamefont
  {de~Rugy}}\ and\ \bibinfo {author} {\bibfnamefont {D.}~\bibnamefont
  {Sternad}},\ }\bibfield  {title} {\enquote {\bibinfo {title} {Interaction
  between discrete and rhythmic movements: reaction time and phase of discrete
  movement initiation during oscillatory movements},}\ }\href@noop {}
  {\bibfield  {journal} {\bibinfo  {journal} {Brain Research}\ }\textbf
  {\bibinfo {volume} {994}},\ \bibinfo {pages} {160--174} (\bibinfo {year}
  {2003})}\BibitemShut {NoStop}%
\bibitem [{\citenamefont {Strogartz}(1994)}]{strogartz1994nonlinear}%
  \BibitemOpen
  \bibfield  {author} {\bibinfo {author} {\bibfnamefont {S.~H.}\ \bibnamefont
  {Strogartz}},\ }\bibfield  {title} {\enquote {\bibinfo {title} {Nonlinear
  dynamics and chaos: With applications to physics, biology},}\ }\href@noop {}
  {\bibfield  {journal} {\bibinfo  {journal} {Chemistry and Engineering}\
  }\textbf {\bibinfo {volume} {441}} (\bibinfo {year} {1994})}\BibitemShut
  {NoStop}%
\bibitem [{\citenamefont {Di~Bernardo}\ \emph {et~al.}(2008)\citenamefont
  {Di~Bernardo}, \citenamefont {Budd}, \citenamefont {Champneys}, \citenamefont
  {Kowalczyk}, \citenamefont {Nordmark}, \citenamefont {Tost},\ and\
  \citenamefont {Piiroinen}}]{di2008bifurcations}%
  \BibitemOpen
  \bibfield  {author} {\bibinfo {author} {\bibfnamefont {M.}~\bibnamefont
  {Di~Bernardo}}, \bibinfo {author} {\bibfnamefont {C.~J.}\ \bibnamefont
  {Budd}}, \bibinfo {author} {\bibfnamefont {A.~R.}\ \bibnamefont {Champneys}},
  \bibinfo {author} {\bibfnamefont {P.}~\bibnamefont {Kowalczyk}}, \bibinfo
  {author} {\bibfnamefont {A.~B.}\ \bibnamefont {Nordmark}}, \bibinfo {author}
  {\bibfnamefont {G.~O.}\ \bibnamefont {Tost}}, \ and\ \bibinfo {author}
  {\bibfnamefont {P.~T.}\ \bibnamefont {Piiroinen}},\ }\bibfield  {title}
  {\enquote {\bibinfo {title} {Bifurcations in nonsmooth dynamical systems},}\
  }\href@noop {} {\bibfield  {journal} {\bibinfo  {journal} {SIAM review}\
  }\textbf {\bibinfo {volume} {50}},\ \bibinfo {pages} {629--701} (\bibinfo
  {year} {2008})}\BibitemShut {NoStop}%
\bibitem [{\citenamefont {Bernardo}\ \emph {et~al.}(2008)\citenamefont
  {Bernardo}, \citenamefont {Budd}, \citenamefont {Champneys},\ and\
  \citenamefont {Kowalczyk}}]{bernardo2008piecewise}%
  \BibitemOpen
  \bibfield  {author} {\bibinfo {author} {\bibfnamefont {M.}~\bibnamefont
  {Bernardo}}, \bibinfo {author} {\bibfnamefont {C.}~\bibnamefont {Budd}},
  \bibinfo {author} {\bibfnamefont {A.~R.}\ \bibnamefont {Champneys}}, \ and\
  \bibinfo {author} {\bibfnamefont {P.}~\bibnamefont {Kowalczyk}},\ }\href@noop
  {} {\emph {\bibinfo {title} {Piecewise-smooth dynamical systems: theory and
  applications}}},\ Vol.\ \bibinfo {volume} {163}\ (\bibinfo  {publisher}
  {Springer Science \& Business Media},\ \bibinfo {year} {2008})\BibitemShut
  {NoStop}%
\bibitem [{\citenamefont {Gaspard}(1990)}]{gaspard1990measurement}%
  \BibitemOpen
  \bibfield  {author} {\bibinfo {author} {\bibfnamefont {P.}~\bibnamefont
  {Gaspard}},\ }\bibfield  {title} {\enquote {\bibinfo {title} {Measurement of
  the instability rate of a far-from-equilibrium steady state at an infinite
  period bifurcation},}\ }\href@noop {} {\bibfield  {journal} {\bibinfo
  {journal} {Journal of Physical Chemistry}\ }\textbf {\bibinfo {volume}
  {94}},\ \bibinfo {pages} {1--3} (\bibinfo {year} {1990})}\BibitemShut
  {NoStop}%
\bibitem [{\citenamefont {Santos}\ \emph {et~al.}(2004)\citenamefont {Santos},
  \citenamefont {Rivera}, \citenamefont {Eiswirth},\ and\ \citenamefont
  {Parmananda}}]{santos2004effects}%
  \BibitemOpen
  \bibfield  {author} {\bibinfo {author} {\bibfnamefont {G.~J.~E.}\
  \bibnamefont {Santos}}, \bibinfo {author} {\bibfnamefont {M.}~\bibnamefont
  {Rivera}}, \bibinfo {author} {\bibfnamefont {M.}~\bibnamefont {Eiswirth}}, \
  and\ \bibinfo {author} {\bibfnamefont {P.}~\bibnamefont {Parmananda}},\
  }\bibfield  {title} {\enquote {\bibinfo {title} {Effects of noise near a
  homoclinic bifurcation in an electrochemical system},}\ }\href@noop {}
  {\bibfield  {journal} {\bibinfo  {journal} {Physical Review E}\ }\textbf
  {\bibinfo {volume} {70}},\ \bibinfo {pages} {021103} (\bibinfo {year}
  {2004})}\BibitemShut {NoStop}%
\bibitem [{\citenamefont {Nurujjaman}, \citenamefont {Iyengar},\ and\
  \citenamefont {Parmananda}(2008)}]{nurujjaman2008noise}%
  \BibitemOpen
  \bibfield  {author} {\bibinfo {author} {\bibfnamefont {M.}~\bibnamefont
  {Nurujjaman}}, \bibinfo {author} {\bibfnamefont {A.~S.}\ \bibnamefont
  {Iyengar}}, \ and\ \bibinfo {author} {\bibfnamefont {P.}~\bibnamefont
  {Parmananda}},\ }\bibfield  {title} {\enquote {\bibinfo {title}
  {Noise-invoked resonances near a homoclinic bifurcation in the glow discharge
  plasma},}\ }\href@noop {} {\bibfield  {journal} {\bibinfo  {journal}
  {Physical Review E}\ }\textbf {\bibinfo {volume} {78}},\ \bibinfo {pages}
  {026406} (\bibinfo {year} {2008})}\BibitemShut {NoStop}%
\bibitem [{\citenamefont {Gon{\c{c}}alves}(2005)}]{gonccalves2005regions}%
  \BibitemOpen
  \bibfield  {author} {\bibinfo {author} {\bibfnamefont {J.~M.}\ \bibnamefont
  {Gon{\c{c}}alves}},\ }\bibfield  {title} {\enquote {\bibinfo {title} {Regions
  of stability for limit cycle oscillations in piecewise linear systems},}\
  }\href@noop {} {\bibfield  {journal} {\bibinfo  {journal} {IEEE Transactions
  on Automatic Control}\ }\textbf {\bibinfo {volume} {50}},\ \bibinfo {pages}
  {1877--1882} (\bibinfo {year} {2005})}\BibitemShut {NoStop}%
\bibitem [{\citenamefont {Hogan}\ and\ \citenamefont
  {Sternad}(2007)}]{hogan2007rhythmic}%
  \BibitemOpen
  \bibfield  {author} {\bibinfo {author} {\bibfnamefont {N.}~\bibnamefont
  {Hogan}}\ and\ \bibinfo {author} {\bibfnamefont {D.}~\bibnamefont
  {Sternad}},\ }\bibfield  {title} {\enquote {\bibinfo {title} {On rhythmic and
  discrete movements: reflections, definitions and implications for motor
  control},}\ }\href@noop {} {\bibfield  {journal} {\bibinfo  {journal}
  {Experimental brain research}\ }\textbf {\bibinfo {volume} {181}},\ \bibinfo
  {pages} {13--30} (\bibinfo {year} {2007})}\BibitemShut {NoStop}%
\bibitem [{\citenamefont {Flash}\ and\ \citenamefont
  {Hogan}(1985)}]{flash1985coordination}%
  \BibitemOpen
  \bibfield  {author} {\bibinfo {author} {\bibfnamefont {T.}~\bibnamefont
  {Flash}}\ and\ \bibinfo {author} {\bibfnamefont {N.}~\bibnamefont {Hogan}},\
  }\bibfield  {title} {\enquote {\bibinfo {title} {The coordination of arm
  movements: an experimentally confirmed mathematical model},}\ }\href@noop {}
  {\bibfield  {journal} {\bibinfo  {journal} {Journal of neuroscience}\
  }\textbf {\bibinfo {volume} {5}},\ \bibinfo {pages} {1688--1703} (\bibinfo
  {year} {1985})}\BibitemShut {NoStop}%
\bibitem [{\citenamefont {Uno}, \citenamefont {Kawato},\ and\ \citenamefont
  {Suzuki}(1989)}]{uno1989formation}%
  \BibitemOpen
  \bibfield  {author} {\bibinfo {author} {\bibfnamefont {Y.}~\bibnamefont
  {Uno}}, \bibinfo {author} {\bibfnamefont {M.}~\bibnamefont {Kawato}}, \ and\
  \bibinfo {author} {\bibfnamefont {R.}~\bibnamefont {Suzuki}},\ }\bibfield
  {title} {\enquote {\bibinfo {title} {Formation and control of optimal
  trajectory in human multijoint arm movement},}\ }\href@noop {} {\bibfield
  {journal} {\bibinfo  {journal} {Biological cybernetics}\ }\textbf {\bibinfo
  {volume} {61}},\ \bibinfo {pages} {89--101} (\bibinfo {year}
  {1989})}\BibitemShut {NoStop}%
\bibitem [{\citenamefont {Todorov}\ and\ \citenamefont
  {Jordan}(2002)}]{todorov2002optimal}%
  \BibitemOpen
  \bibfield  {author} {\bibinfo {author} {\bibfnamefont {E.}~\bibnamefont
  {Todorov}}\ and\ \bibinfo {author} {\bibfnamefont {M.~I.}\ \bibnamefont
  {Jordan}},\ }\bibfield  {title} {\enquote {\bibinfo {title} {Optimal feedback
  control as a theory of motor coordination},}\ }\href@noop {} {\bibfield
  {journal} {\bibinfo  {journal} {Nature neuroscience}\ }\textbf {\bibinfo
  {volume} {5}},\ \bibinfo {pages} {1226--1235} (\bibinfo {year}
  {2002})}\BibitemShut {NoStop}%
\bibitem [{\citenamefont {Ikegami}\ \emph {et~al.}(2010)\citenamefont
  {Ikegami}, \citenamefont {Hirashima}, \citenamefont {Taga},\ and\
  \citenamefont {Nozaki}}]{ikegami2010asymmetric}%
  \BibitemOpen
  \bibfield  {author} {\bibinfo {author} {\bibfnamefont {T.}~\bibnamefont
  {Ikegami}}, \bibinfo {author} {\bibfnamefont {M.}~\bibnamefont {Hirashima}},
  \bibinfo {author} {\bibfnamefont {G.}~\bibnamefont {Taga}}, \ and\ \bibinfo
  {author} {\bibfnamefont {D.}~\bibnamefont {Nozaki}},\ }\bibfield  {title}
  {\enquote {\bibinfo {title} {Asymmetric transfer of visuomotor learning
  between discrete and rhythmic movements},}\ }\href@noop {} {\bibfield
  {journal} {\bibinfo  {journal} {Journal of Neuroscience}\ }\textbf {\bibinfo
  {volume} {30}},\ \bibinfo {pages} {4515--4521} (\bibinfo {year}
  {2010})}\BibitemShut {NoStop}%
\bibitem [{\citenamefont {d'Avella}, \citenamefont {Saltiel},\ and\
  \citenamefont {Bizzi}(2003)}]{d2003combinations}%
  \BibitemOpen
  \bibfield  {author} {\bibinfo {author} {\bibfnamefont {A.}~\bibnamefont
  {d'Avella}}, \bibinfo {author} {\bibfnamefont {P.}~\bibnamefont {Saltiel}}, \
  and\ \bibinfo {author} {\bibfnamefont {E.}~\bibnamefont {Bizzi}},\ }\bibfield
   {title} {\enquote {\bibinfo {title} {Combinations of muscle synergies in the
  construction of a natural motor behavior},}\ }\href@noop {} {\bibfield
  {journal} {\bibinfo  {journal} {Nature neuroscience}\ }\textbf {\bibinfo
  {volume} {6}},\ \bibinfo {pages} {300--308} (\bibinfo {year}
  {2003})}\BibitemShut {NoStop}%
\bibitem [{\citenamefont {Takei}\ \emph {et~al.}(2017)\citenamefont {Takei},
  \citenamefont {Confais}, \citenamefont {Tomatsu}, \citenamefont {Oya},\ and\
  \citenamefont {Seki}}]{takei2017neural}%
  \BibitemOpen
  \bibfield  {author} {\bibinfo {author} {\bibfnamefont {T.}~\bibnamefont
  {Takei}}, \bibinfo {author} {\bibfnamefont {J.}~\bibnamefont {Confais}},
  \bibinfo {author} {\bibfnamefont {S.}~\bibnamefont {Tomatsu}}, \bibinfo
  {author} {\bibfnamefont {T.}~\bibnamefont {Oya}}, \ and\ \bibinfo {author}
  {\bibfnamefont {K.}~\bibnamefont {Seki}},\ }\bibfield  {title} {\enquote
  {\bibinfo {title} {Neural basis for hand muscle synergies in the primate
  spinal cord},}\ }\href@noop {} {\bibfield  {journal} {\bibinfo  {journal}
  {Proceedings of the National Academy of Sciences}\ }\textbf {\bibinfo
  {volume} {114}},\ \bibinfo {pages} {8643--8648} (\bibinfo {year}
  {2017})}\BibitemShut {NoStop}%
\bibitem [{\citenamefont {Degallier}\ \emph {et~al.}(2011)\citenamefont
  {Degallier}, \citenamefont {Righetti}, \citenamefont {Gay},\ and\
  \citenamefont {Ijspeert}}]{degallier2011toward}%
  \BibitemOpen
  \bibfield  {author} {\bibinfo {author} {\bibfnamefont {S.}~\bibnamefont
  {Degallier}}, \bibinfo {author} {\bibfnamefont {L.}~\bibnamefont {Righetti}},
  \bibinfo {author} {\bibfnamefont {S.}~\bibnamefont {Gay}}, \ and\ \bibinfo
  {author} {\bibfnamefont {A.}~\bibnamefont {Ijspeert}},\ }\bibfield  {title}
  {\enquote {\bibinfo {title} {Toward simple control for complex, autonomous
  robotic applications: combining discrete and rhythmic motor primitives},}\
  }\href@noop {} {\bibfield  {journal} {\bibinfo  {journal} {Autonomous
  robots}\ }\textbf {\bibinfo {volume} {31}},\ \bibinfo {pages} {155--181}
  (\bibinfo {year} {2011})}\BibitemShut {NoStop}%
\bibitem [{\citenamefont {Schuster}\ and\ \citenamefont
  {Marhl}(2001)}]{schuster2001bifurcation}%
  \BibitemOpen
  \bibfield  {author} {\bibinfo {author} {\bibfnamefont {S.}~\bibnamefont
  {Schuster}}\ and\ \bibinfo {author} {\bibfnamefont {M.}~\bibnamefont
  {Marhl}},\ }\bibfield  {title} {\enquote {\bibinfo {title} {Bifurcation
  analysis of calcium oscillations: time-scale separation, canards, and
  frequency lowering},}\ }\href@noop {} {\bibfield  {journal} {\bibinfo
  {journal} {Journal of Biological Systems}\ }\textbf {\bibinfo {volume} {9}},\
  \bibinfo {pages} {291--314} (\bibinfo {year} {2001})}\BibitemShut {NoStop}%
\end{thebibliography}%

\end{document}